\documentclass[AMA,STIX2COL]{WileyNJD-v2} 

\usepackage{moreverb}
\usepackage{xcolor}
\usepackage{lipsum}
\usepackage{subcaption}

\newcommand\BibTeX{{\rmfamily B\kern-.05em \textsc{i\kern-.025em b}\kern-.08em
T\kern-.1667em\lower.7ex\hbox{E}\kern-.125emX}}

\articletype{Feature Article}%

\received{<day> <Month>, <year>}
\revised{<day> <Month>, <year>}
\accepted{<day> <Month>, <year>}

\begin{document}

\title{PECVD and PEALD on polymer substrates (part I): Fundamentals and analysis of plasma activation and thin film growth}

\author[1]{Teresa de los Arcos}
\author[2]{Peter Awakowicz}
\author[3]{Jan Benedikt}
\author[4]{Beatrix Biskup}
\author[4]{Marc B\"oke}
\author[5,6]{Nils Boysen}
\author[4]{Rahel Buschhaus}
\author[7]{Rainer Dahlmann}
\author[5,6]{Anjana Devi}
\author[8]{Tobias Gergs}
\author[2]{Jonathan Jenderny}
\author[4]{Achim von Keudell}
\author[9]{Thomas D. K\"uhne}
\author[7]{Simon Kusmierz}
\author[1]{Hendrik M\"uller}
\author[8]{Thomas Mussenbrock}
\author[10]{Jan Trieschmann}
\author[5]{David Zanders}
\author[9]{Frederik Zysk}
\author[1]{Guido Grundmeier*}

\address[1]{\orgdiv{Technical and Macromolecular Chemistry}, \orgname{Paderborn University}, \orgaddress{Warburger Str. 100, 33098 Paderborn, \country{Germany}}}

\address[2]{\orgdiv{Chair of Electrical Engineering and Plasma Technology}, \orgname{Ruhr University Bochum}, \orgaddress{Universitaetsstrasse 150, 44801 Bochum, \country{Germany}}}

\address[3]{\orgdiv{Institute of Experimental and Applied Physics}, \orgname{Christian-Albrechts-University Kiel}, \orgaddress{Leibnizstraße 11-19, 24098 Kiel, \country{Germany}}}

\address[4]{\orgdiv{Chair of Experimental Physics II: Physics of Reactive Plasmas}, \orgname{Ruhr University Bochum}, \orgaddress{Universitaetsstrasse 150, 44801 Bochum, \country{Germany}}}

\address[5]{\orgdiv{Inorganic Materials Chemistry}, \orgname{Ruhr University Bochum}, \orgaddress{Universitaetsstrasse 150, 44801 Bochum, \country{Germany}}}

\address[6]{\orgname{Fraunhofer IMS}, \orgaddress{47057 Duisburg, Germany}}

\address[7]{\orgdiv{Institute for Plastics Processing}, \orgname{RWTH Aachen University}, \orgaddress{Seffenter Weg 201, 52074 Aachen, \country{Germany}}}

\address[8]{\orgdiv{Chair of Electrodynamics and Plasma Technology}, \orgname{Ruhr University Bochum}, \orgaddress{Universitaetsstrasse 150, 44801 Bochum, \country{Germany}}}

\address[9]{\orgdiv{Chair of Theoretical Chemistry}, \orgname{Paderborn University}, \orgaddress{Warburger Str. 100, 33098 Paderborn, \country{Germany}}}

\address[10]{\orgdiv{Theoretical Electrical Engineering}, \orgname{Kiel University}, \orgaddress{Kaiserstraße 2, 24143 Kiel, \country{Germany}}}

\authormark{G. Grundmeier \textsc{et al}}

\corres{*G. Grundmeier, Technical and Macromolecular Chemistry, Paderborn University, 33098 Paderborn, Germany\\ \email{guido.grundmeier@uni-paderborn.de}}

\abstract[Abstract]{This feature article presents recent results on the analysis of plasma/polymer interactions and the nucleation of ultra-thin plasma films on polymeric substrates. Because of their high importance for the understanding of such processes, in-situ analytical approaches of the plasma volume as well as the plasma/substrate interfaces are introduced prior to the findings on plasma surface chemistry. The plasma activation of polymeric substrates is divided in the understanding of fundamental processes on model substrates and the relevance of polymer surface complexity.
Concerning thin film nucleation and growth both PECVD and PEALD processes as well as the combination of both processes are considered both for model substrates and technical polymers. 
Based on the comprehensive presentation of recent results the selective perspectives of this research field are discussed
}


\maketitle

\section{Introduction} \label{sec1}


This feature article considers the analysis of the initial states of film growth on polymer substrates. The assembled results are based on the cooperation between research groups in the field of plasma physics, chemistry, electric as well as mechanical engineering over the last years, mostly within the frame of the transregional project SFB-TR 87 ("Pulsed high power plasmas for the synthesis of nanostructured functional layers"). This feature article aims at bridging the gap between the understanding of plasma processes in the gas phase and the resulting surface and interface processes of the polymer. The results show that interfacial adhesion and initial film growth can be well controlled and even predicted based on the combination of analytical approaches.

\subsection{Applications, relevance of the topic}\label{subsec11}

There exists a wide range of approaches to deposit thin barrier films in order to overcome the high permeation rate of plastics in packing applications \cite{Lange.2003,Chatham.1996,ZHANG.2018,Vartiainen.2014,LIU.2022,Czeremuszkin.2001}. Dry deposition techniques range from plasma enhanced chemical vapor deposition (PECVD), or magnetron sputtering to atomic layer deposition (ALD) as well as combinations of them. 

Such barrier layers can be divided into inorganic films such as silicon oxides (SiO$_x$), aluminum oxide (Al$_2$O$_3$), zinc oxide (ZnO) or titanium oxide (TiO$_2$), and organic films such as amorphous, hydrogenated carbon (a-C:H) and tetrahedral amorphous carbon (ta-C). The barrier improvement factor (BIF) of these coatings varies over several orders of magnitude and also depends on the substrate material \cite{Plog.2011,Schneider.2007}. 

However, the application of plasma thin film technologies and plasma enhanced ALD in membrane applications is only rarely considered \cite{Bryjak.2010,Jiang.2014,Nagasawa.2013,Wolden.2017,Kleines.2020}. Nevertheless, due to the inherent stability of highly crosslinked PECVD and PEALD films such films could be of relevance for application, in which the long-term stability of the film is an issue.

\subsection{The role of plasma surface activation}\label{subsec12}

Plasma activation has been extensively studied as a way to create polar groups on low-energy polymer surfaces such as polyethylene or perfluorinated polymers \cite{Liston.1993}.  
Plasma based surface activation of polymers both at low or ambient pressure is of crucial importance for the functionality of the polymeric part or as an intermediate step prior to film deposition or adhesive joining. 
The plasma-induced surface reactions create in many cases oxygen containing groups such as carboxylic acid, carbonyls, alcohol, or amino groups which lead to an increased wettability by fluids, promote the formation of continuous and well adhering thin PVD films or enable the formation of interfacial chemical bonds between adhesives or organic coatings \cite{Jaritz.2017}.
Moreover, the deposition of thin films by means of PECVD and PEALD has been reported to be very sensitive to the prior plasma activation step. Many polymeric substrates require a plasma activation in order to achieve high bonding strength to the deposited PECVD or PEALD films \cite{Hoppe.2022}.

The general effects that a plasma process has on polymer surfaces and surface-near regions are relatively well known. The combination of characterization techniques such as X-ray photoelectron spectroscopy (XPS), time of flight secondary ion mass spectrometry (ToF-SIMS) or Fourier transform infrared spectroscoy (FTIR) has shown how plasma processes lead to the scission of chemical bonds within the polymer; a process that is followed by the saturation of dangling bonds with e.g. terminating oxygen containing groups \cite{Hoppe.2022}. 
The complexity of the plasma/surface interaction is due to the combination of a multitude of active species present in plasma and the parallel UV-light exposure,  which combined often lead to a variety of surface functional groups \cite{Jaritz.2017}. Therefore, control over the surface chemistry can only be achieved once there is a fundamental understanding of the correlation between plasma parameters and the plasma-induced reactions in the surface-near region of the polymer. 

A generally accepted model describing the state of the polymer surface under plasma treatment was illustrated by Corbella \textit{et al}. \cite{Corbella.2015}. A two-fold modified top layer is formed due to the different penetration ranges of the different species present in the plasma. Neutrals and ions at intermediate energies have very limited penetration depth of the order of 2–3 nm; by comparison, UV radiation from the glow discharge penetrates the polymer by tens of nanometers. Thus, neutrals and ions mainly cause etching due to chain scission and induce roughness changes and cross-linking in the first monolayers. In the deeper region, photons are involved in chain scission followed by polymer cross-linking.
However, only recently very fundamental in-situ studies aiming at separation between different fundamental plasma-surface interaction mechanisms 
have been published by Corbella \textit{et al} \cite{Corbella.2019}.

\subsection{Challenges regarding the understanding and process control}\label{subsec13}


Both in the case of PECVD and PEALD the interfacial reactions occurring at the polymer surface due to the interaction with the plasma (independently of whether the plasma is used as a pretreatment/activation tool or as a deposition tool) are challenging to analyze and to describe on a molecular level. The plasma phase is highly complex and it interacts with a substrate that is both reactive to the plasma and at the same time shows a high mobility of macromolecular fragments created by plasma-induced chain scission. Plasma-activated polymer surfaces tend to react with the environment, which strongly limits the conclusions drawn from ex-situ analytical studies. Furthermore, even in the case of in-situ analytical studies such as optical reflection spectroscopy, the gradual differences in the dielectric properties between the substrate and the deposited film hamper the application of interface and thin film sensitive analytical techniques. Only few approaches to date combined the analysis of the plasma phase with in-situ interface analytical techniques. As presented in this feature article the study of model polymeric and molecular surfaces in combination with a defined state of the plasma phase is a key to the understanding of technically relevant and thereby complex systems. 

\subsection{Scope of the article}\label{subsec14}

The aim of this feature article is to present the most recent state of knowledge regarding the analysis and understanding of plasma-induced polymer surface activation and its effect on the subsequent nucleation and growth of PECVD and PEALD thin films. Moreover, the analysis and quantification of relevant plasma parameters is shown to provide a basis for the prediction of the resulting surface chemical processes at the plasma/polymer interface.


\section{In-situ analytical approaches}\label{sec2}

A fundamental understanding of the processes at the plasma polymer interface can be reached at best, if in-situ real time data are available. Due to the low pressure of a plasma environment, optical diagnostics or sampling via pumping interfaces are  used, which are in most cases rather easy to implement although the analysis of the collected data might be challenging. This has to be complemented with various ex-situ diagnostics to characterize the plasma exposed surfaces.

\subsection{Plasma volume (transport and reactions)}\label{subsec21}


In order to achieve a high degree of process control, in-situ monitoring of process parameters is essential. In this regard the gas phase chemistry is of central importance, involving plasma parameters such as electron density and electron temperature, fluxes of ions, neutrals and radicals as well as the type of species formed for deposition. Depending on the desired parameters to be studied, various diagnostic methods are available.

\begin{figure}[htb]
\begin{center}
\includegraphics[width=0.49\textwidth]{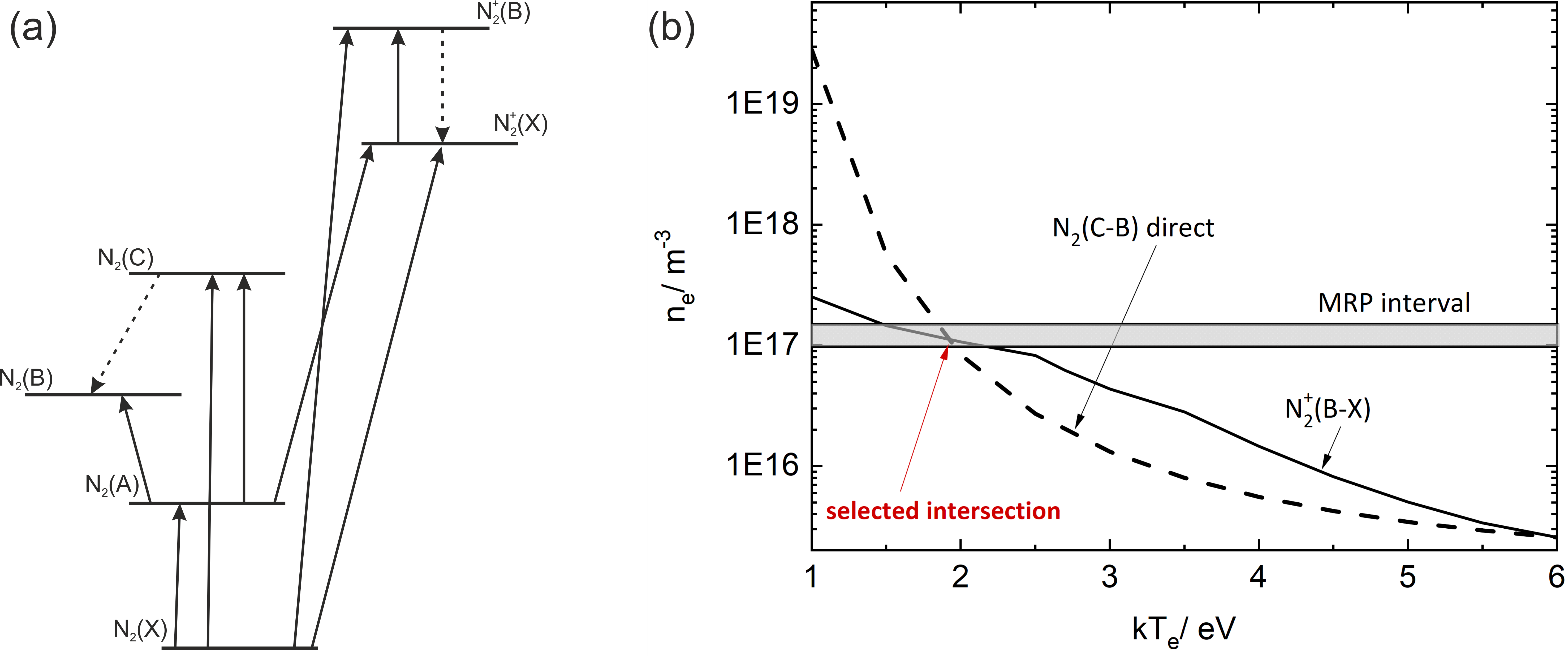}
\end{center}
\caption{\textbf{(a)} Nitrogen collisional radiative model used to calculate electron density and electron temperature. \textbf{(b)} N$_2$(C-B) and N$_2^+$(B-X) emission lines in dependence of electron density and electron temperature as obtained from OES measurements. Selection of the intersection is possible due to additional MRP measurements. Adapted with permission from Steves \textit{et al.} \cite{Steves.2013}. Copyright 2013 IOP Publishing.} \label{fig:OES-CRModel_MRP}
\end{figure}

Allowing for the non-invasive determination of both plasma parameters and chemical composition of the gas phase, optical methods offer a wide range of applications. When absolutely calibrated and equipped with a suitable collisional radiative model, optical emission spectroscopy (OES) allows for the determination of gas temperature, electron temperature and electron density as well as densities of reactive species \cite{Bibinov.2007, Steves.2013}. (Since OES needs only optical access, possible disturbances of the process are avoided, although this comes at a cost of losing spatial resolution due to line of sight signal integration.) As an example, Steves \textit{et al}. \cite{Steves.2013} used OES in combination with a N$_2$ collisional radiative model  to calculate the atomic oxygen density in the deposition of silicon oxide barrier coatings. The collisional radiative model considered nitrogen emission of the N$_2$(C-B) and the N$_2^+$(B-X) transitions. These were assumed to be excited by direct electron-impact excitation of ground states N$_2$(X) and N$_2^+$(X) and by stepwise electron-impact excitation via metastable state N$_2$(A). The complete set of states included in the collisional radiative model are displayed in figure \ref{fig:OES-CRModel_MRP} (a). Comparison of the two emission lines in dependency of electron density and electron temperature yields the measured electron density and electron temperature at the intersecting points. In case of more than one intersection point, a priori knowledge on the observed process or additional information is needed for selection of the correct intersection point. The profile of both emission lines and the subsequent selection of an intersection point is shown in figure \ref{fig:OES-CRModel_MRP} (b) for an exemplary measurement from Steves et al. \cite{Steves.2013} additionally using a multipole resonance probe (MRP) for comparison of the results from OES.

The method was then used to compare the influence of an RF discharge superimposing a pulsed microwave discharge in oxygen. Results indicated a high electron density for the microwave discharge and a low electron density for the RF discharge. In consequence, adding the RF discharge to the microwave discharge did not lead to a significant increase of electron density during the microwave pulse. Instead, the role of the RF discharge can be explained by increasing the atomic oxygen fluence during deposition as atomic oxygen is still produced during the pulse pause. This was confirmed by Mitschker \textit{et al}. \cite{Mitschker.2018} stating the RF discharge leads to increased incorporated ion energy during microwave discharge and to increased production of atomic oxygen during pulse pause.

In addition to determination of electron density and electron temperature, Steves \textit{et al}. \cite{Steves.2013} used the obtained emission spectrum to calculate the density of atomic oxygen during deposition considering dissociative excitation and direct electron impact excitation. Knowledge of the atomic oxygen density is crucial for deposition of barrier coatings as this is one of the key driving factors for obtaining dense, highly cross-linked layers inducing a significant improvement in barrier properties towards oxygen \cite{Mitschker.2018}.

Further insight into fragmentation patterns and gas phase chemistry during the deposition process can be gained by laser based spectroscopy. Nave \textit{et al}. demonstrated the detection of absolute densities of deposition products and excess gases by applying infrared laser absorption spectroscopy (IRLAS) \cite{Nave.2016}.

An alternative approach to characterise the gas phase chemistry of a deposition process is the application of a mass spectrometer. Depending on the configuration, a mass spectrometer can provide insight into the fragmentation pattern, detection of film-forming species and by-products, radical densities and ion energy distribution functions. When equipped with a multi-channel scalar card, temporal resolution down to the microseconds becomes possible \cite{Benedikt.2009}. Determination of radical densities is often difficult due to their short lifetimes and thereby resulting low densities. Benedikt \textit{et al}. \cite{Benedikt.2005} showed that by performing threshold ionisation mass spectrometry (TIMS) such low radical densities can successfully be resolved. Wavhal \textit{et al}. \cite{Wavhal.2006} used mass spectrometry to gain insight into the fragmentation of hexamethyldisiloxane (HMDSO) diluted in oxygen for deposition of SiO$_2$ coatings. The study revealed Si$_2$O(CH$_3$)$_5^+$ with a mass of 147\,u to be the dominant dissociation channel during fragmentation due to the loss of one methyl group. In combination with a signal at 32\,u corresponding to the molecular oxygen, production of different silicon-containing fragments was observed. Depending on parameters such as dilution rate of HMDSO and the discharge power, different degrees of fragmentation were reported. Increasing either discharge power or oxygen content lead to a reduction of carbon content in the films deposited, which is confirmed by a decreasing signal of Si$_2$O(CH$_3$)$_5^+$ and an increased signal of volatile products such as CO and H$_2$O. In terms of film quality this corresponds to a change from polymer-like SiOCH films to inorganic SiO$_x$ films. A detailed overview on the fragmentation pattern of HMDSO is given by Blanchard \textit{et al}. \cite{Blanchard.2015}.

In addition to experimental diagnostics (or when lacking access for diagnostics), simulations can greatly benefit the process understanding and development. Applications of such simulations are based in the fundamental understanding of the plasma process in terms of heating and excitation \cite{Kushner.2009,Donko.2012,Schulze.2018}, the gas flow distribution \cite{Kirchheim.2018} or in analysis of the gas phase chemistry describing the reactions taking place during plasma ignition and deposition \cite{Gudmundsson.1999, Kemaneci.2015}. Depending on the type of model embedded, spatial and time resolved information can be gained.

Kirchheim \textit{et al}. demonstrated that gas flow simulations limited to the non-reactive gas dynamics can be exploited to optimize the deposition of silicon oxide SiO$_x$ coatings from hexamethyldisiloxane and oxygen (HMDSO/O$_2$) precursor gases. \cite{Kirchheim.2018} Specifically, a pulsed microwave plasma-enhanced chemical vapor deposition (MW-PECVD) with substrates of up to 300 mm by 300 mm and a 2.45 GHz plasmaline array was investigated. A particle-based transport simulation based on the open-source simulation software OpenFOAM\cite{wellerOpenFOAMFoundationLtd2016} was used for the prediction of the precursor gas density distribution in front of a substrate. In a model-based design approach, the geometry and distribution of the gas inlets were optimized. The final design constituted of a remote index plate with gas inlet for O$_2$, paired with a lance matrix gas inlet for HMDSO. The simulation approach suggested a significant improvement in precursor gas homogeneity. A model-based redesign of the experimental setup verified a similar improvement for the deposited SiO$_x$ coatings residual stress, coating thickness, and water vapor transmission rate.

 Kemaneci \textit{et al} \cite{Kemaneci.2019} developed a global model to investigate the chemical compounds formed in the gas phase for another silicon oxide deposition process involving hexamethyldisiloxane (HMDSO) diluted in oxygen. The model included over 1000 reactions and allowed for calculation of electron density, electron temperature, and densities of a variety of compounds formed during deposition. The model is therefore well suited to complement or even replace diagnostics such as the OES or mass spectrometry characterization discussed earlier in this section. Comparison of measurement values to simulation results using this model showed good agreement in electron density and electron temperature for a variation of discharge power, process pressure and oxygen to HMDSO ratio. Neutral densities such as for HMDSO, CO or CH$_4$ were also in good agreement to measurements from Nave \textit{et al}. \cite{Nave.2016}. Using the model, the authors were additionally able to gain insight into the underlying reaction mechanisms. They found only little formation of negative O$^-$ ions confirming earlier assumptions by Steves \textit{et al}. \cite{Steves.2013} and the conversion of methyl initially dissociated from HMDSO being converted into various types of hydrocarbons. In terms of radical species, atomic oxygen and hydrogen fluxes were stated to be responsible for removal of carbon and hydrocarbon species on the surfaces thereby being an important factor in production of low carbon-containing silicon oxide coatings. 


\subsection{Plasma/polymer interfaces}\label{subsec22}
\label{ch:2.2}


As for the case of plasma analysis, the in-situ analysis of the plasma-polymer interface relies mostly on optical methods, compatible with the pressure environment required for the plasma. Thus, spectroscopic reflectometry or ellipsometry can be used to monitor the optical constants of a thin film system as it develops during film growth or etching. This method is monolayer sensitive, so that minute changes in composition and film thickness can be resolved.

However, techniques such as reflectometry and ellipsometry have the disadvantage that the analysis requires the definition of optical models for the measurement process that require the knowledge of optical constants, specific surface roughness and a value for the film thickness (or a set of film thicknesses for multilayer systems). This is usually an ill-posed problem, since the determination of an optical constant for example for a complex optical system is not unique. Therefore, these methods needs to be complemented by other ex-situ or in-situ analysis tools such as profilometry, photoelectron spectroscopy or infrared spectroscopy. 

In infrared spectroscopy, the change in the absorption of chemical groups is followed by monitoring the absorption of vibrational transitions. This is straightforward and contains detailed information on the prevalent binding in the film. Infrared spectroscopy is usually less sensitive than optical spectroscopy in the visible range, but by exploiting multiple reflection techniques nanometer sensitivity can be reached. Typical examples are the use of an external reflection geometry such as in infrared reflection absorption spectroscopy (IRRAS) \cite{Grundmeier.2015} or an optical cavity substrate \cite{vonKeudell.2002}. With the adequate substrate preparation (i.e. preparation of thin polymer layers onto highly reflective metal surfaces) it is also possible to perform in-situ IRRAS analysis of the effect of plasma treatment on the polymer surface \cite{Xie.2022} or to follow the early stages of nucleation when the polymer is to be coated by PECVD techniques \cite{Dietrich.2018, Hoppe.2017}.

Photoelectron spectroscopy is a method widely used for the investigation of chemical changes at surfaces. The relatively small information depth associated to the technique (typically below 10 nm) makes it uniquely appropriate for the investigation of plasma-induced surface modifications, particularly when the analysis can be performed in vacuum, without exposure of the sample to air \cite{Ozkaya.2014}.


\section{Plasma activation of polymeric substrates}\label{sec3}

The interaction of a plasma with a polymeric substrate is at the core of the method of plasma-induced surface functionalization or the deposition of a good quality coatings onto a polymer. A plasma provides a cocktail of species that can affect the surface stoichiometry and morphology depending on the nature of the species and their penetration depth. This interaction, however, is not unidirectional, because also the polymer may contain a multitude of volatile species that desorb into the plasma and thereby affect the plasma properties. Therefore, it is important to assess the impact of the plasma on the polymer, but also the impact of the polymer on the plasma.  



\subsection{Fundamental aspects of plasma/polymer interactions in PP, PET and PC}\label{subsec31}


The surface modification of a polymer by a plasma involves two steps. At first an activation or cleaning step of the polymer interface leading to a change in wettability and/or the stoichiometry of the topmost layer. In some cases, a second step of thin film deposition follows to create for example a gas barrier layer on top. Especially, the first step prior to any film deposition is crucial, because the system of polymer and coating can fail due to improper adhesion at the interface. The elementary processes during plasma treatment are governed by the impact of ions, electrons, radicals, metastables and photons on the top most layers. To study this impact, we regard the case of an argon plasma with oxygen that is used to modify the topmost layer of a polymer.

The effects of the different species from the plasma are very different and depend on their typical penetration depth. Electrons and radicals are stopped within the very first topmost layers, whereas ions may penetrate a few nanometer into the solid depending on their ion energy. Photons, however, may penetrate even further and can reach depths of the order of the photon wavelength. As a result, the polymer toplayer may be crosslinked by the VUV photons over a few 100 nanometer, whereas the stoichiometry of the very top layer is affected by the impact of ions and radicals.

\begin{figure}[htb]
\begin{center}
\includegraphics[width=0.5\textwidth]{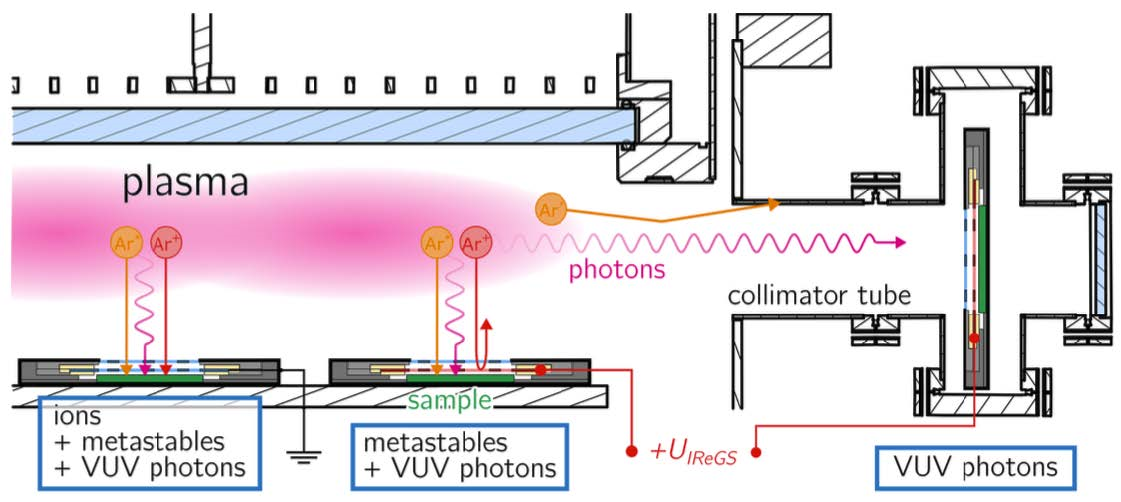}
\end{center}
\caption{Separation of plasma species in an ICP discharge by an Ion Repelling Grid System (IReGS) for three cases: \textbf{i)} IReGS on ground: all species reach the substrate (left), \textbf{ii)} repelling grid is biased and no ions reach the substrate (middle), \textbf{iii)} IReGS is positioned behind a collimator and almost exclusively photons reach the substrate (right). Here, the treatment time is adjusted to the same photon flux as in i) and ii). 
\label{fig:iregspolymer}}
\end{figure}

These elementary aspects of plasma polymer interaction were explored in a inductively coupled plasma (ICP) experiment and in a beam experiment with a focus on separating the impact of the different species: (i) In the ICP \textit{plasma experiment} \cite{Schlebrowski.2013}, a grid system was used to repel the ions from impacting the polymer surface and to study thereby the impact of photons and reactive neutral species with and without ion impact (for details of the ion-repelling grid system (IReGS) see \cite{Biskup.2018}). The impact of photons only was studied by placing the polymer samples behind a space filter to trap all heavy species such as ions, metastables and radicals from reaching the sample surface. This is illustrated in figure \ref{fig:iregspolymer}. (ii) In the beam experiment \cite{Budde.2018,Corbella.2015,Grosse-Kreul.2013,Grosse-Kreul.2013a,vonKeudell.2017}, quantified beams of argon ions and oxygen atoms were send to polymer samples and the impact on the film surface was monitored via infrared absorption spectroscopy\cite{Ozkaya.2014}. The ion source employed an electron cyclotron resonance (ECR) plasma for ionization and created an energy selected ion beam. Two sets of experiments were performed, one with direct line-of-sight from the ECR source to the sample where the VUV photons from the ECR plasma can reach the sample and another using a bending optic to block the VUV photons but to steer the plasma beam onto the sample. 

In the ICP plasma experiment, the polymers polypropylene (PP) and polyethylene terephthalate (PET) were treated with a pulsed argon plasma at a power of 500 W, a duty cycle of 20\%, and a pressure of 3 Pa. The plasma conditions and plasma parameters are similar to the ones in \cite{Bahre.2013, Behm.2014}. The pulsing mitigates any plasma induced heating of the samples. The PP and PET foils are exposed to the plasma for different time spans and the surface energies are measured using contact angle measurements with the liquids water and diiodomethane. Thereby, the polar and non-polar contributions and their modification to the surface energy can be separated. 

\begin{figure}[htb]
\begin{center}
\includegraphics[width=0.5\textwidth]{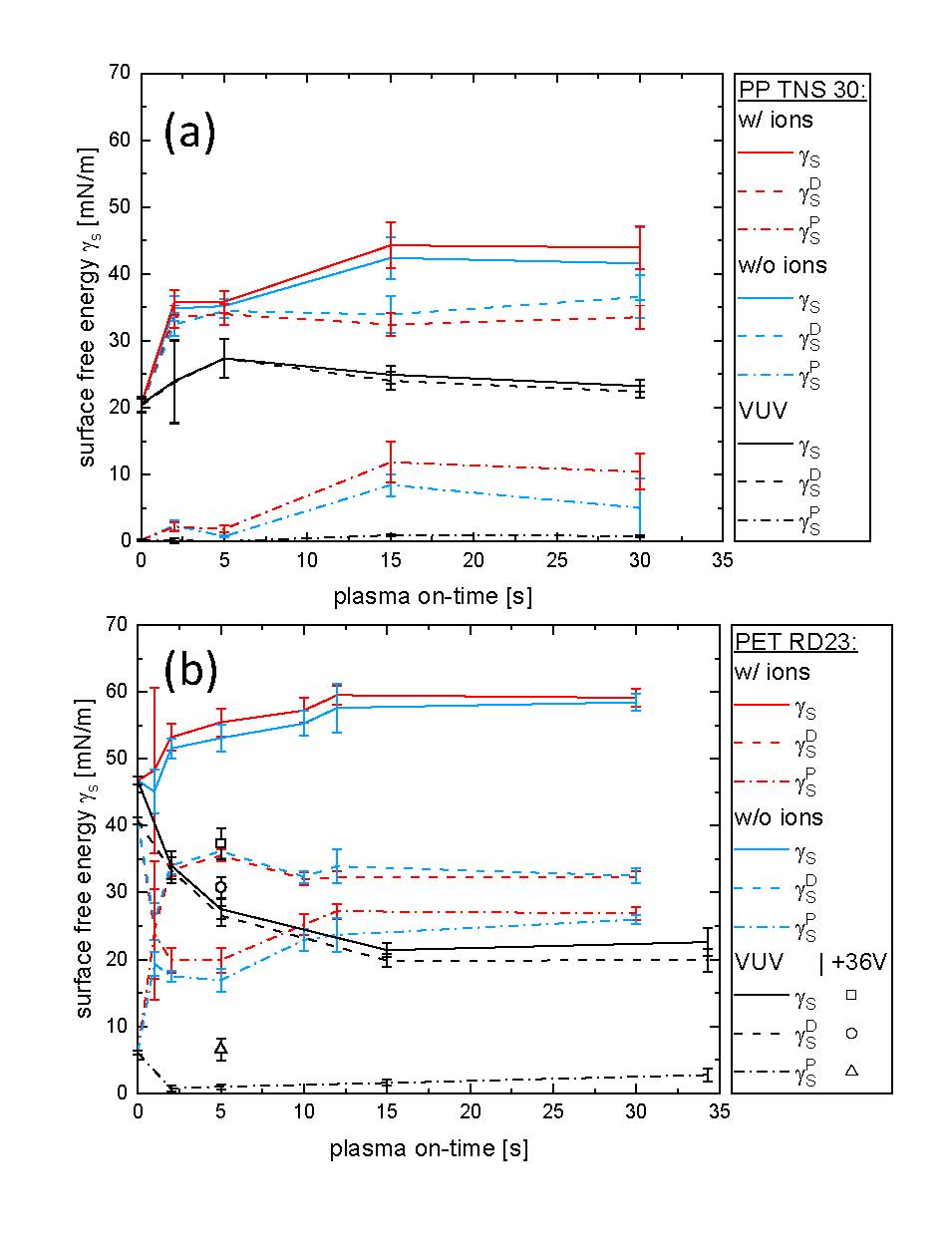}
\end{center}
\caption{Surface free energy $\gamma_s$ development for PP \textbf{(a)} and PET \textbf{(b)} with increasing treatment time. The polar $\gamma_s^P$ and dispersive part $\gamma_s^D$ is shown as well.} \label{fig:iregspolymerPPPET}
\end{figure}

Figure \ref{fig:iregspolymerPPPET} shows the change in surface energy separated in the polar ($\gamma_s^P$) and non-polar ($\gamma_s^D$) part for the polymers PP (a) and PET (b) for different plasma treatment times. The impact of the plasma treatment with and without ions showed an increase in surface energy for PP and for PET. The comparison of the experiment with and without ions does not show a significant difference. This increase in surface energy originates mainly from the non-polar part for PP in the very first minutes of plasma exposure before the change in the contribution from the polar part dominates at longer exposure time. This is in contrast to PET, where the non-polar part stayed rather constant and the increase in surface energy was governed by an increase in the polar part. This increase is much faster for PET compared to PP.  

Experiments with exposure to only VUV photons show that the surface energy in the case of PP was affected only to a small degree, whereas the surface energy significantly decreased for PET. This photon-only induced decrease in surface energy, however, was completely compensated by the increase in surface energy during plasma treatment.

\begin{figure}[htb]
\begin{center}
\includegraphics[width=0.5\textwidth]{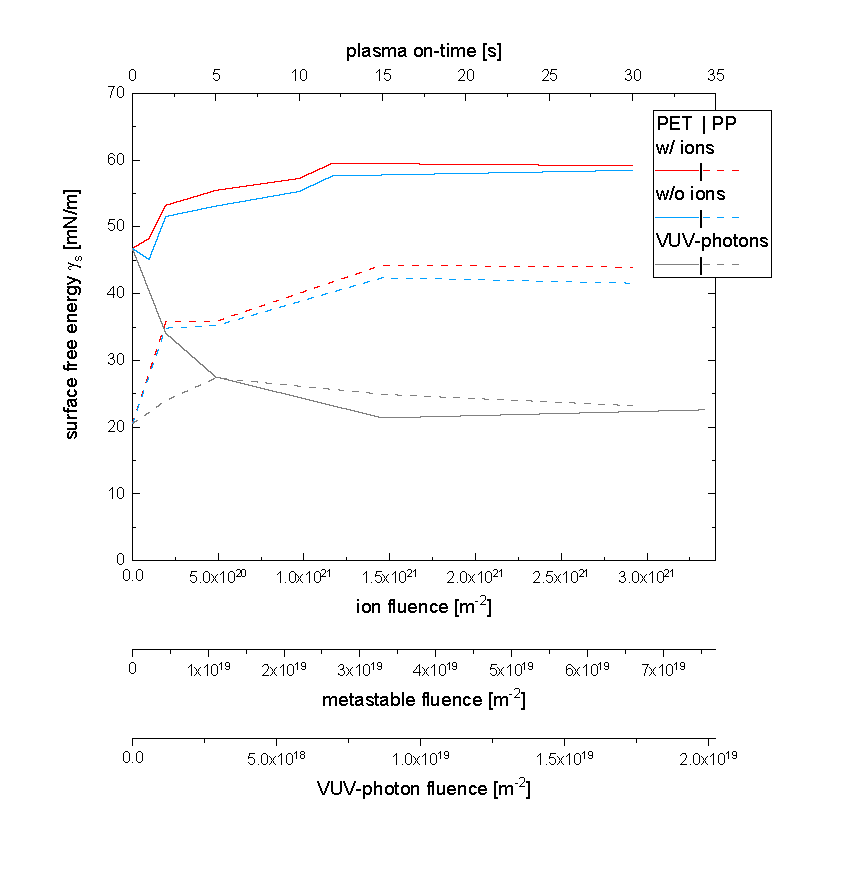}
\end{center}
\caption{Surface energy of PET and PP substrates in comparison with the particle fluence. The \textbf{red data} shows the surface energy influence by the whole plasma cocktail including metastables, ions and photons, while the \textbf{blue data} is the influence without ions and the grey data the influence solely by photons. \label{fig:iregsPPPETQuantitative}}
\end{figure}

The plasma exposure was quantified by measuring absolute fluxes of VUV photons, argon ions and argon metastables (for details about the methods see \cite{Behm.2014, Schroter.2014}). The fluences reached after 30 s of plasma exposure were 3 $\cdot$ 10$^{21}$ m$^{-2}$ for the ions, 7 $\cdot$ 10$^{19}$ m$^{-2}$ for metastables, and 2 $\cdot$ 10$^{19}$ m$^{-2}$ for VUV photons. These fluences are shown as additional scales in figure \ref{fig:iregsPPPETQuantitative} summarizing the changes in surface energy for PET and PP.

One can see that the impact of reactive species from the argon plasma causes an increase in surface energy. It is interesting to note that the comparison of experiments with and without ions show almost no difference although ions constitute the largest fraction of reactive species in the incident flux. It should be kept in mind, however, that the surface energy is sensitive to the change in the very first monolayer, so that ions penetrating deeper into the material can be less efficient than metastable species which recombine directly 
at the surface. A typical surface coverage of a monolayer constitutes 10$^{19}$ m$^{-2}$, which coincides with the fluence that is reached by the metastables, when the change in surface energy starts to saturate. Apparently, the changes in surface energy can be linked to the impact of metastables at the very surface.

The impact of VUV photons on PET showed a decrease in surface energy, whereas PP seemed to be unaffected. The VUV photons penetrate deep into the material and  may cause chain scission and chain crosslinking. PET contains oxygen containing groups in the material, whereas PP contains only hydrocarbon groups. If we assume that VUV photons cause chain scission within the PET chains, these become more mobile and the polar groups rotate into the film since this corresponds to a lower surface energy. This is similar to the known aging of a surface that had been functionalized to a hydrophilic state. The polymer PP is already non polar and exhibits a low surface energy, so no photon-induced changes are expected. 

This leads to the question, how argon metastables are able to induce the formation of polar groups at the surface although argon is an inert gas. We postulate that residual water vapor either from the background of the plasma reactor, adsorbed to the polymer sample beforehand or from the ambient after venting the plasma chamber may adsorb on the topmost layer at dangling bond sites that are created by metastables during plasma treatment. Since the contact angle measurements are sensitive to only the top most surface, very small concentrations of water vapor are sufficient to reach full coverage. 

During the plasma experiments only the change in surface energy was monitored being induced by a pure argon plasma and a possible reaction with water vapor during or after a very long plasma treatment time. Such treatment times could be significantly accelerated by the intentional use of oxygen in a plasma process where oxygen atom fluxes are usually comparable to the ion fluxes, so that a surface functionalization could be reached in a fraction of a second rather than 30 seconds. 
Such a contribution of oxygen, however, may cause chemical sputtering leading to an erosion of the sample. Nevertheless, the plasma on-time can be kept very short, so that the net erosion remains small but an efficient incorporation of polar groups is realized. 

The impact of chemical sputtering on polymer treatment is studied for the case of polycarbonate (PC) in the beam experiment employing argon ions at 500 eV and oxygen atoms from a thermal particle source. The VUV photon flux has also been quantified to be 2.15 $\cdot$ 10$^{16}$ m$^{-2}$s$^{-1}$ by using a sodium silicylate film to convert the VUV photons into visible light that is measured, as detailed in \cite{Iglesias.2017,Bibinov.2007}. The O atom flux is 4 $\cdot$ 10$^{17}$ m$^{-2}$s$^{-1}$. Two sets of experiments can be formed using the ion source with and without an ion bending unit to block the direct line-of-sight from the plasma to the sample. The ion flux is quantified at an energy of 500 eV yielding 1.2 $\cdot$ 10$^{18}$ m$^{-2}$s$^{-1}$ without the bending unit and 1 $\cdot$ 10$^{17}$ m$^{-2}$s$^{-1}$ with the bending unit in place. This shows that the ion flux and the photon flux in the beam experiment are very similar to the plasma experiment above.  For details see \cite{Budde.2018}.

\begin{figure}[htb]
\begin{center}
\includegraphics[width=0.5\textwidth]{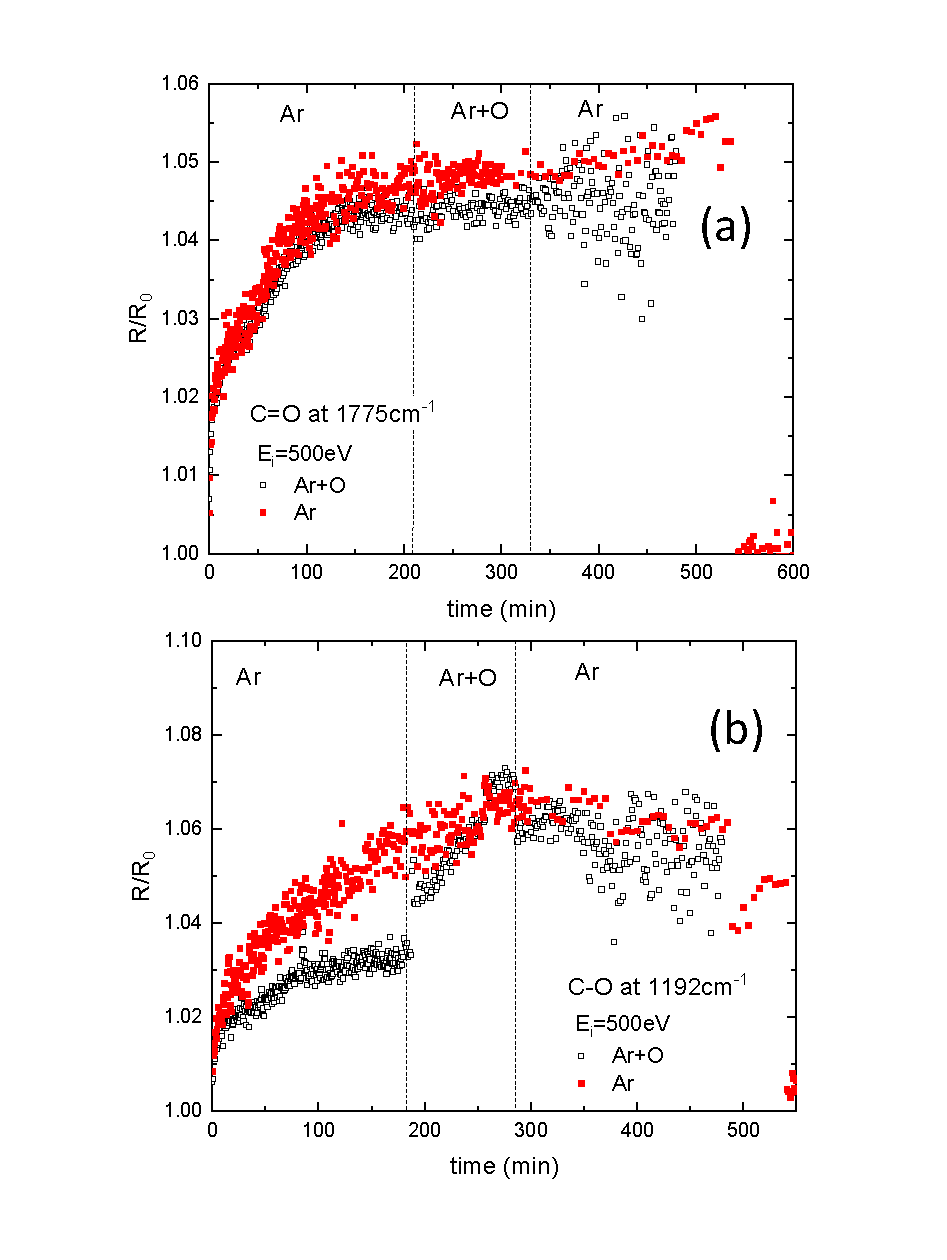}
\end{center}
\caption{Change in reflectivity $R/R_0$ due to the loss in CO \textbf{(a)} and C=O groups \textbf{(b)} in polycarbonate during exposure to argon ions at 500 eV and oxygen atoms \textit{without} VUV radiation.  \label{fig:pcbeam1}}
\end{figure}

PC already contains CO groups that can be monitored with infrared spectroscopy via monitoring the vibrations at 1775 cm$^{-1}$ for C=O groups and at 1192 cm$^{-1}$ for C-O groups. The reflectivity spectra ratio of measurement to background $R/R_0$ is presented.
Figure \ref{fig:pcbeam1} shows the first experiment with the VUV photons being blocked. The O atom source is switched on after approximately 210 min for one hour as indicated by dashed lines in figure \ref{fig:pcbeam1}. One can see that after the onset of the particle beams, the reflectivity $R/R_0$ increases significantly due to the loss of C=O groups and of C-O groups. This is due to sputtering of the polymer surface. This loss, however, saturates after 100 to 150 min, because we assume that the continuous ion bombardment causes a cross linking of the surface layer, which makes it less susceptible to etching. After 210 minutes, the oxygen atom beam is switched on, and the concentration of CO is strongly decreasing, as can be seen from an increase in reflectivity, whereas the concentration of C=O groups seems to be unaffected. The addition of oxygen, apparently supports the more efficient removal of C-O groups in the PC polymer. The C=O groups constitute end groups attached to the side of the PC polymer chain, whereas the C-O groups are parts directly within the PC polymer chain. Therefore, the impact of O atoms may correspond to an additional contribution to break the polymer chain, although the complete removal is not affected as can be seen from the lack in any change in the C=O signal. After switching off the O atom source, the C-O signal slowly recovers indicating that the C-O groups reform. This may occur due to a reaction of the plasma activated polymer with residual oxygen in the ambient or by a relaxation of the polymer leading to a re-connection of a broken polymer chains.

\begin{figure}[htb]
\begin{center}
\includegraphics[width=0.5\textwidth]{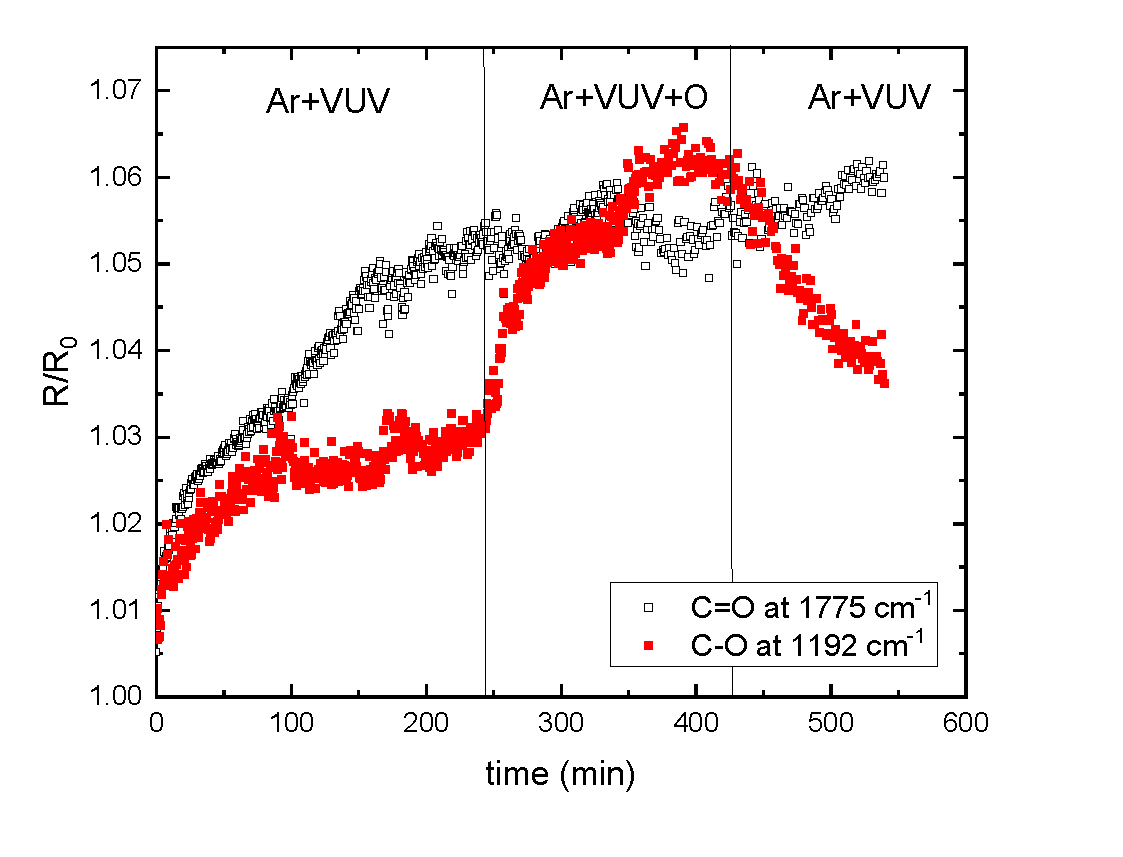}
\end{center}
\caption{Change in reflectivity $R/R_0$ due to the loss in CO and C=O groups in polycarbonate during exposure to argon ions at 500 eV and oxygen atoms \textit{with} VUV radiation. \label{fig:pcbeam2}}
\end{figure}

A similar trend can also be observed in the experiment without blocking the VUV photons as shown in figure \ref{fig:pcbeam2}. Also here, the O atom beam is switched on after 210 minutes showing a significant loss of C-O groups although the number of C=O groups is not being affected. The total removal of C-O groups with and without the presence of VUV photons is rather similar. 

It is interesting to note that the concentration of C-O groups recovers after the O atom flux is switched off. This could be explained, if we assume that the VUV photons cross link the polymer which becomes visible as a creating of new C-O groups in the PC polymer chain. The source of oxygen might be O$_2$ or H$_2$O in the chamber background. This recovery of the C-O groups after switching off the O beam source is much stronger when VUV photons are present. This could be either due to a stronger activation of the surface and thus a more efficient uptake of O from the ambient, or a photon-induced acceleration of the polymer chain cross linking.

Summarizing, one can state that the plasma impact of an oxygen containing argon plasma causes a very quick change in the surface composition, where the erosion by chemical sputtering causes mainly a removal of C-O groups consistent ion induced breaking of the PC polymer chains. It is expected that the surface is polar in nature.     

\begin{figure}[htb]
\begin{center}
\includegraphics[width=0.5\textwidth]{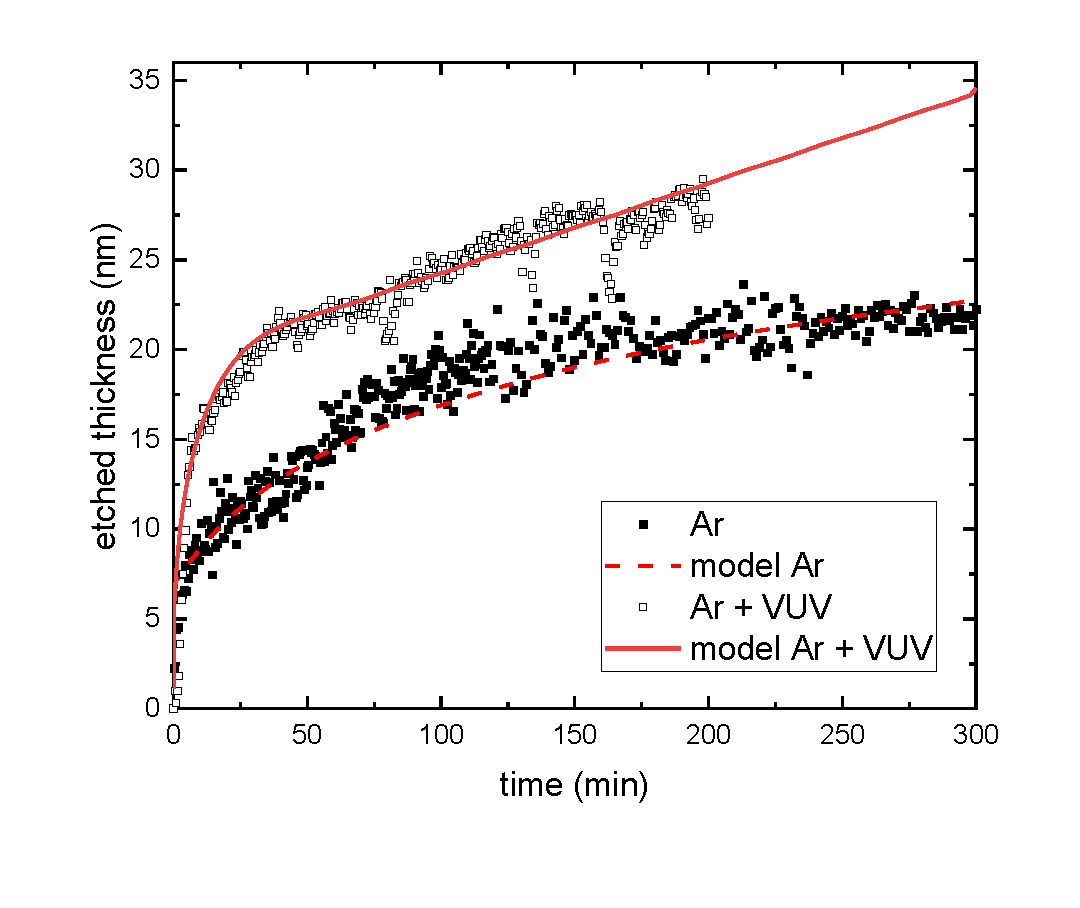}
\end{center}
\caption{Net etching of an initial 30nm polycarbonate film by the argon ion beam at 500 eV with and without the impact of VUV photons. The dashed and solid lines denote a rate equation modeling as detailed in the text. \label{fig:pcbeam3}}
\end{figure}

Finally, the influence of photons on the net etch rate by argon ions has been analyzed by following the change in the reflectivity $R/R_0$ at 1775 cm$^{-1}$ corresponding to C=O groups and calibrating that change to the signal after complete removal of the initial 30 nm spin coated PC film. Thereby, the change in reflectivity can be converted to a reduction in film thickness, as shown in figure \ref{fig:pcbeam3} for etching of PC by argon ions only, but with and without the presence of VUV photons. One can see that the initial etching is very fast and a slower etch rate establishes after a removal of 15 nm of film. Apparently, the ion bombardment depletes the surface of hydrogen and oxygen leaving a surface which can no longer so easily be attacked by ion impact. The additional VUV radiation has, at first glance, only a small impact on this etching dynamics. This is analyzed in more detail in the following.

Figure \ref{fig:pcbeam3} also shows a simple rate equation modeling (solid and dashed lines) \cite{Budde.2018} to calculate the etched film thickness from a balance equation for the coverage $\Theta$ of ion-induced cross linked surface sites versus pristine polymer sites:

\begin{equation}
n_0 \frac{d\theta}{dt} = j_{ions} Y_a (1-\theta) - j_{ions} Y_c\theta
\end{equation}

with the surface site density $n_0$, the probability $Y_a$ for crosslinking a pristine polymer site and $Y_c$ for sputtering a cross-linked site and exposing a pristine site, which is calculated to $Y_c$= 0.03 from SRIM simulations for 500 eV argon ion physical sputtering of a C:H=1:1 amorphous film. The net sputter rate $ER$ is then:

\begin{equation}
ER = j_{ions}Y_c\theta + j_{ions}Y_p(1-\theta)
\end{equation}
with $Y_p$ the sputter yield for pristine sites surface sites. The fitted values are listed in table \ref{tab:sputteringpc}.
\begin{table}[h]
    \centering
    \begin{tabular}{|c|c|c|}\hline
     parameter  &  with VUV & without VUV \\ \hline
        $Y_c$ &  0.03 & 0.03 \\
        $Y_p$ & 0.92 & 12  \\
        $Y_a$ & 0.06 & 0.05 \\\hline
    \end{tabular}
    \caption{Parameters for the rate equation modeling of the sputtering of polycarbonate.}
    \label{tab:sputteringpc}
\end{table}

 The modeled curves fit very well the data. The fitted sputter yields for the pristine polymer $Y_p$  are much larger than the physical sputter yield $Y_c$. Such large values can be explained by assuming that a single bond breaking event in a longer polymer chain may release larger fragments that desorb. More interesting is the observation that the presence of VUV light reduces the etch yield $Y_p$ for the pristine polymer from 12 to 0.92. We assume that the VUV light continuously cross-links the polymer in depths beyond the penetration range of incident ions. As a result, the ions interact with an VUV-induced cross-linked polymer rather than a pristine polymer. Consequently, the polymer chains are much smaller leading to smaller fragment and thus smaller yields. The yield for the ion-induced conversion of a pristine site $Y_a$ is around 0.05 irrespective of the presence of VUV light and similar to the physical sputter yield $Y_c$. 

Summarizing, plasma experiments on PP and PET showed an activation of the polymer surface by an increase in surface energy. Ions and VUV photons play a minor role since their impact is dominant in the bulk material. In PP a change of the non-polar part is followed by a dominant contribution from the polar part at longer exposure time, whereas in PET the increase in surface energy is governed by an increase in the polar part. This increase is much faster for PET compared to PP.  The exposure of PC to VUV, as investigated in the beam experiment, leads to a strong cross-linking of the surface and a reduced sputter yield of the pristine polymer.

\subsection{Plasma activation - relevance of substrate composition}\label{subsec32}



The surface activation of polymers, or plastics by means of plasma processes covers a very broad range of different polymeric materials, ranging from polyolefines to biopolymers. Taking into account that most technical polymers contain a variety of additives and even fillers, it is clear that the complex composition of the substrate will strongly affect the plasma activation process. These processes have already been described in several excellent reviews covering a broad variety of polymers \cite{Ghobeira.2022,Kusano.2014,Williams.2013}. The aim of the present work is to illustrate the correlation of polymer surface composition with the plasma activation based on selected examples relevant for packaging applications and gas membranes.




In comparison to high surface energy materials, such as metals and glasses, polymers have a considerably lower tendency to promote nucleation and thin film growth due to their lower surface energy.  However, this aspect can be improved by plasma treatment i.e., plasma activation of the surface increase bonding between the substrates surface and the applied thin film. Within the scope of the presented work plasma activation of the commonly used polyethylene terephthalate (PET), polypropylene (PP) representing the group of plastics for packaging and the membrane material polydimethyl siloxane will be discussed. 

Plasma activation of PET is typically performed by exposing the surface to an oxygen plasma which leads to introduction of functional groups onto the surface of the polymer. These functional groups mainly consist of polar groups such as hydroxyl (-OH), carboxyl (-COOH) and carbonyl (-C=O) groups. Incorporation of these groups increases the surface energy of the PET leading to improved surface properties, e.g. increased wettability and film adhesion. XPS measurements of oxygen plasma treated PET surfaces show an increase of carbon bound in ester groups (-O-C=O) in respect to aromatic bound carbon (-CAr) for pretreatment times of 1s and 10s. \cite{Bahre_2013}

\begin{figure}[htb]
\begin{center}
\includegraphics[width=0.5\textwidth]{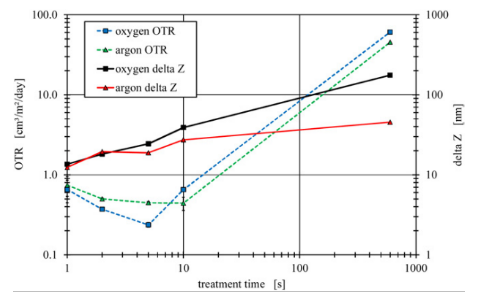}
\end{center}
\caption{Height differences for oxygen and argon pretreated PET film and resulting OTR with an applied SiO$_x$ thin film of 30 nm thickness over treatment time for oxygen and argon plasmas. Adapted with permission from \cite{Bahre_2013}. Copyright 2013 IOP Publishing.
}
\label{fig:ikv2}
\end{figure}

Additionally, to the modification due to the functional groups, exposure to an oxygen or argon plasma increases the surface roughness of the PET. This etching effect results in an adhesion improvement of subsequent coating. Since the height difference increases exponentially with the treatment time the formation of a coherent barrier coating in the subsequent coating process with a SiO$_x$ barrier thin film is compromised resulting in rising oxygen transmission rates (OTR), as illustrated in figure \ref{fig:ikv2}\cite{Bahre_2013}. This is due to uneven ablation of the PET during the treatment as well as the exposition of anti-block particles.  Therefore, the pretreatment time has to be chosen carefully to avoid overtreatment. \cite{Behm.2014} The indicated pretreatment time in figure \ref{fig:ikv2} corresponds to an oxygen fluence in the order of $10^{19}$ - $10^{20}~m^{-2}$.

Polyolefins, such as polyethylene (PE) and polypropylene (PP), are partially crystalline non-polar polymers used in a wide range of molecular weights and applications. Due to their considerably lower surface energy compared to polyesters such as PET, polyolefins must be prepared differently to ensure adhesion between the PECVD silicon based thin films. Polypropylene (PP) consists mainly of C-C and C-H with bond energies of approx. 3.6 eV and 4.3 eV, respectively. Achieving covalent bonds within the interface is desirable thus breaking the polymer bonds is necessary which can be done by e.g., ion bombardment or photon radiation. For photon radiation, especially vacuum ultraviolet radiation (VUV) within the range of 200 nm to 50 nm is able to break those bonds i.e., creating scissions within the PP backbone structure. While the VUV emission from oxygen plasmas has major role in surface activation of PP, considerably less UV radiation is emitted from argon plasmas.\cite{Jaritz_2017}  Since plasma treatment of polyolefin surfaces results in breaking of bonds within the polymer chains a certain degradation of the surface is unavoidable. Nevertheless, overtreatment can result in formation of a layer of low molecular weight oxidized polymer (LWMOP) on the surface which reduces the bond strength due to lower cohesion within the LWMOP layer.\cite{Jaritz_2017} 

\begin{figure}[htb]
\begin{center}
\includegraphics[width=0.5\textwidth]{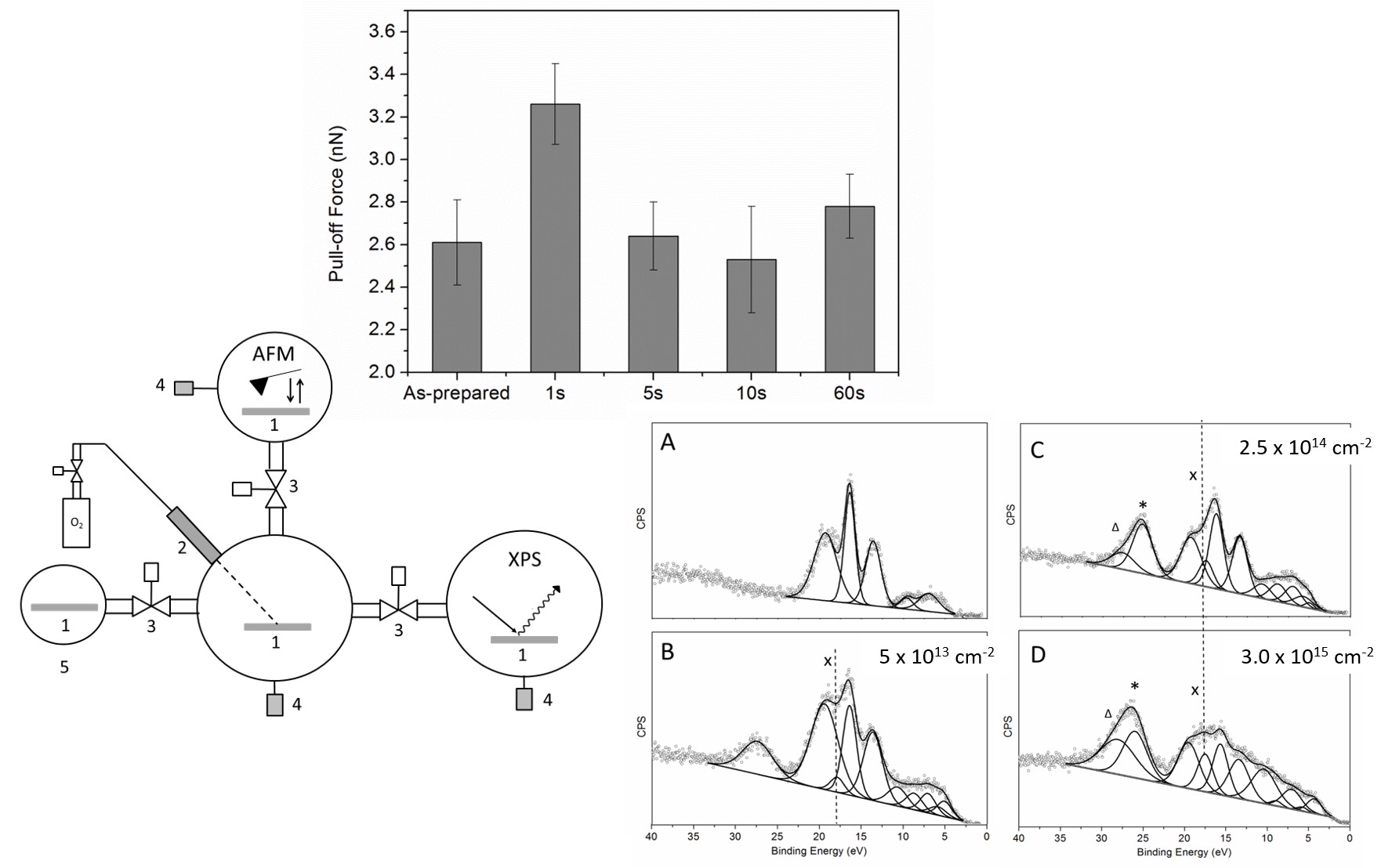}
\end{center}
\caption{Overview of the combination of in-vacuo approaches using XPS for surface analysis and AFM based chemical force microscopy for the correlation of adhesive properties with plasma surface activation. \textbf{Bottom left:} Schematic of the UHV cluster for in situ XPS-VB, and chemical force microscopy measurements. \textbf{Bottom right:} X-ray valence band spectra of \textbf{(A)} as prepared, \textbf{(B)} 1 s, \textbf{(C)} 5 s, and \textbf{(D)} 60 s oxygen plasma-treated polypropylene surfaces. \textbf{Top:} AFM-based chemical force microscopy measurements in UHV with SiO$_2$–tip. Adapted with permission from Ozkaya \textit{et al.} \cite{Ozkaya.2014}. Copyright 2014 Wiley and Sons.} \label{fig:grundmeier32}
\end{figure}

Oezkaya \textit{et al}. analysed the plasma modification of polypropylene (PP) and correlated the surface composition with the adhesive properties based on a combination of in-vacuo UHV XPS and AFM studies \cite{Ozkaya.2014} (see Figure \ref{fig:grundmeier32}). Model PP surfaces were exposed to an electron cyclotron resonance (ECR) oxygen plasma. Without the exposure to atmospheric conditions, X-ray excited valence band (VB) spectroscopy and core level X-ray photoelectron spectroscopy (XPS) revealed the changes in the oxidation state and electronic properties. Adhesive properties were analyzed by means of UHV chemical force microscopy. The correlation of the surface spectroscopic data and the analysis of contact forces showed that interactions between a SiO$_2$ and oxygen containing groups in the PP surface are dominated by hydrogen bonds. Such interactions are maximized in the initial phase of surface oxidation while an extended plasma modification leads to the formation of a weak boundary layer.

Unlike polyesters, such as PET, polyolefins do need the application of an intermittent adhesion layer of SiOCH to bond the SiO$_x$ barrier thin film to the surface. Since the adhesion between SiO$_x$ and SiOCH layers is considered to be considerably higher than the bond strength between SiOCH and the polymer surface, the properties of the latter interface must be taken into account.  

\begin{figure}[htb]
\begin{center}
\includegraphics[width=0.5\textwidth]{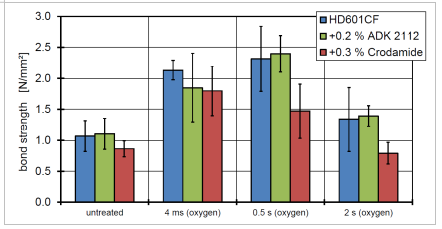}
\end{center}
\caption{Bond strength between SiOCH thin film and pure PP (HD601CF), PP with anti-oxidant (ADK2112) and PP with lubricant (Crodamide) for different pre treatment times. Reprinted from \cite{behm2013plasma} with permission from the authors.
\label{fig:ikv3}}
\end{figure}

In technical applications additives, such as lubricants, antioxidants, UV stabilizers etc., are used to tailor the properties of PP for processing purposes and mechanical properties of the final product. Thus, for industrial application of plasma pretreatments the existence of these additives must be taken into account. Amide-based lubricants, for example, migrate to the surface during PP processing. Therefore, a weak boundary layer is formed and subsequently applied SiOCH coatings have a lower adhesion onto those surfaces. Generally, the adhesion can be improved by short (<0.5s) plasma activation but declines for longer pretreatment times, however, oxidation inhibitors added to the PP do not have a significant impact on the bond strength between PP and SiOCH intermediate layer (figure \ref{fig:ikv3}).\cite{behm2013plasma}

Polymethyldisiloxane (PDMS) can be used as a material for e.g., gas membrane separation. Investigations of the separation properties of PECVD coated multilayer membranes show that the selectivitiy for e.g. CO$_2$/N$_2$ mixtures could be improved.\cite{Kleines_2020} Due to its silicon organic structure and smooth surface PDMS is an ideal substrate for thin film coating oxidates the PDMS surface and leads to a conversion layer of SiO$_x$. The thickness of the conversion layer increases linearly with the pretreatment time and reaches approx. 10 nm after 60s. FTIR measurements of the plasma oxidized PDMS indicate a high crosslinking density within the conversion layer but relatively large pore sizes due to the porous PDMS structure. However, the cumulative void volume could be decreased by oxidative plasma pretreatment. The oxidized conversion layer improves the growth of subsequently applied SiO$_x$ thin films resulting in higher film densities and less overall void volume.\cite{Hoppe.2022} 

\section{Nucleation and thin film growth}\label{sec4}
\subsection{PECVD}\label{subsec41}

Plasma enhanced chemical vapor deposition (PECVD) is a well-known process to deposit thin films on various substrates. It combines cost effective deposition and high deposition rates with a high degree of control over the film characteristics through variation of the plasma parameters. Nevertheless, the control of the initial stages of film growth is crucial, because for, e.g. the plasma deposition of SiO$_x$ on polymer substrates is comprised of concurrent etching and deposition processes, as discussed in detail in the previous paragraphs. Apart from supplying the precursors for the SiO$_x$ film growth, other species typically present in the plasma comprise fast neutrals, metastable atoms, radicals, electrons, ions and (VUV) photons, which can etch or modify in some way the original polymer surface. The consequence is that volatile organic species etched from the surface can intermix with the precursor gas, leading to the formation of weak boundary layers of undetermined composition at the interface. This effect can eventually result in (i) delamination of the SiO$_x$ film, or (ii) to the development of SiO$_x$ films with undesired chemical composition and/or internal structure. Theses aspects of degradation during thin film growth is discussed in more detail in \ref{subsubsec422}.  After the formation of a closed, protective SiO$_x$ layer between the polymer and the plasma, the etching rate of the polymer substrate is negligible and a continuous film growth takes place.

\subsection{PEALD}\label{subsec42}


Unlike the continuous growth characteristic of PECVD, thin film growth in PEALD distinguishes itself by relying on self-limited reactions of the employed precursors with the substrate surfaces and by its cyclic operational mode in which the precursor pulse is separated from the plasma pulse through purge steps with an inert gas (Figure 11, left schematic) \cite{Profijt.2011, Knoops.2019}. These process characteristics lead to smaller growth rates compared to PECVD, but typically facilitate higher thickness- and uniformity control of the deposited films. This is the case for planar substrates such as e.g. dense polymer foils, and also specifically for substrates with three-dimensional features such as porous polymers or membranes \cite{Wolden.2017, Ji.2019}.

\begin{figure}[htb]
\begin{center}
\includegraphics[width=0.5\textwidth]{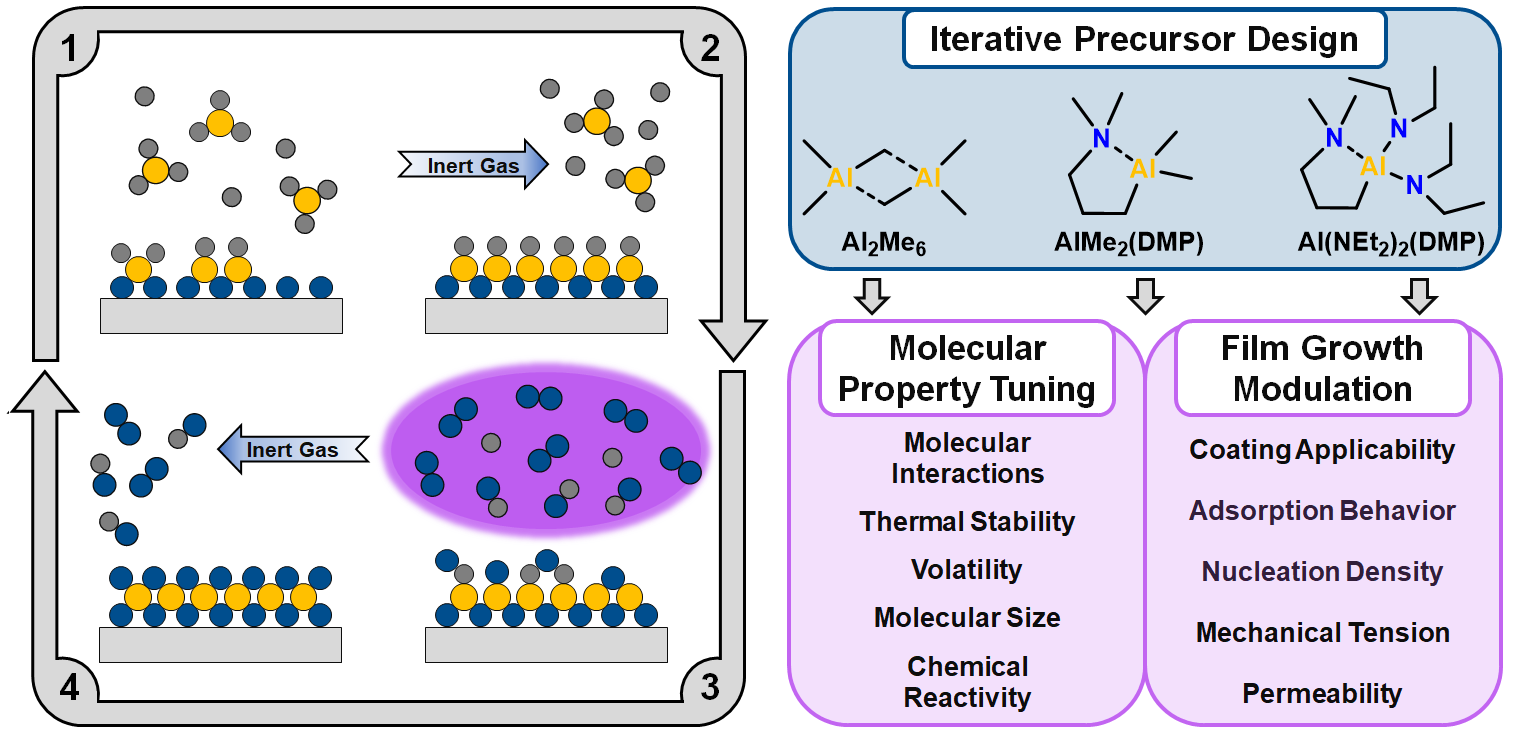}
\end{center}
\caption{\textbf{Left:} Survey on a typical PEALD process sequence consisting of \textbf{1} the precursor pulse, \textbf{2} an inert gas purge, \textbf{3} the plasma pulse and \textbf{4} another purge sequence. \textbf{Right:} The impact of precursor design on molecular properties of the precursor species and implications of these adjusted properties on PEALD film growth and deposit properties.} \label{fig:pealdSchematic}
\end{figure}

Like PECVD, PEALD has been successfully commercialized in the field of encapsulation technology, namely for the deposition of gas or water vapor barrier layers required in highly sensitive organic light emitting diode (OLED) displays \cite{Lee.2018, KOVACS.2021, Lu.2021}. Due to the inherently strong contribution of surface chemistry to film growth in PEALD, it was found that not only the adjustment of plasma parameters in the process allowed to fine-tune thin film properties and thus gas and vapor barrier performances \cite{Arts.2021a, Arts.2021b, Arts.2019, Arts.2020}, but that the applied precursor chemistry can play a pivotal role as well \cite{Mai.2020}. \\
In this context, Mai and co-workers demonstrated that iterative precursor design can unfold in synergetic effects such as the reduction of hazard potential originating from the precursor combined with the preservation or even improvement of thin film barrier properties \cite{Mai.2017, Mai.2019}. They showed that the highly pyrophoric, dimeric Al precursor trimethyl aluminum (TMA, Al$_2$Me$_6$) for the PEALD of aluminum oxide (Al$_2$O$_3$) barrier layers can be efficiently replaced by non-pyrophoric, intramolecularily stabilized alternatives such as [3-(dimethylamino)propyl]dimethyl aluminum (DMAD, AlMe$_2$(DMP)) or bis-(diethylamido)[3-(dimethylamino)propyl] aluminum (BDEADA, Al(NEt$_2$)$_2$(DMP)) (see figure \ref{fig:pealdSchematic}, right schematic).\\
In a related study Gebhard and colleagues showed that the choice of precursor can impact the stress being present in the Al$_2$O$_3$ barrier layers. While Al$_2$O$_3$ layers grown from TMA exhibited tensile stress, those grown from DMAD displayed compressive stress which could be taken advantage of in the context of the gas barrier layer application \cite{Gebhard.2017a}.

\subsection{Role of substrate composition and complexity}\label{subsubsec422}


Similar as in the case of plasma activation, the substrate composition and complexity of the polymeric substrates strongly influences the nucleation and thin film growth by PECVD and PEALD processes.

As already mentioned, the interplay between etching and deposition processes characteristic of plasma-based deposition methods is crucial, with their main impact lying in the early stages of film growth. The evolution of these early stages will determine later on the characteristics of the film in terms of porosity, transmission, or adhesion.

\textbf{Molecular understanding of interface processes:} The analysis of the first nucleation stages can be done very effectively by using well defined model substrates comprised of self assembling monolayers (SAMs). These model substrates can simulate surface changes of specific polymer terminating groups. The plasma-induced changes  in the model surface chemistry can then be visualized using in-situ characterization methods such as polarization modulated infrared reflection absorption spectroscopy (PM-IRRAS). This approach ensures the understanding of the interface processes at a molecular level.

Among the different effects that can lead to the modification of the interfacial chemistry, reactive oxygen species produced by the plasma process can have a great influence. In the case of aliphatic polymeric substrates, these reactive oxygen species can lead to oxidative degradation. Thus for example, Ozkaya \textit{et al}. \cite{Ozkaya.2015} observed the degradation of octadecanethiol self assembling monolayers (ODT-SAMs) when covered with oxygen rich silicon oxide thin layers. These silicon oxide layers were deposited using hexamethyldisiloxane (HMDSO) as precursor diluted in oxygen. Subsequent infrared spectra show the decline of ODT-SAM corresponding peaks in the region between 2800\,$\mathrm{cm^{-1}}$ and 3050\,$\mathrm{cm^{-1}}$ when silicon oxide layers are deposited directly on the ODT-SAMs. In contrast, deposition of an organic adhesion promoting coating, deposited from a purely HMDSO driven process does not lead to deterioration of the ODT-SAMs' peaks. The characteristic peak structure is also maintained when the silicon oxide coating is applied after the adhesion promoting interlayer. 


Further investigations regarding the influence of the atomic oxygen produced during film deposition on the substrate have been conducted by Mitschker \textit{et al}. \cite{Mitschker.2015}. The authors used different O$_2$-HMDSO ratios thereby varying the atomic oxygen fluence the substrate is exposed to during deposition. In addition to the decay of the ODT-SAM characteristic peaks as reported by Ozkaya \textit{et al}. \cite{Ozkaya.2015}, a shift of the Si-O-Si (LO) peak towards higher wavenumbers was observed for increasing atomic oxygen fluence. More importantly, the study revealed the polymeric surface stays intact for low oxygen admixtures. This is important for up-scaling of the process in industrial applications, as a process with only moderately diluted HMDSO exhibits higher deposition rates.

\begin{figure}[htb]
\begin{center}
\includegraphics[width=0.5\textwidth]{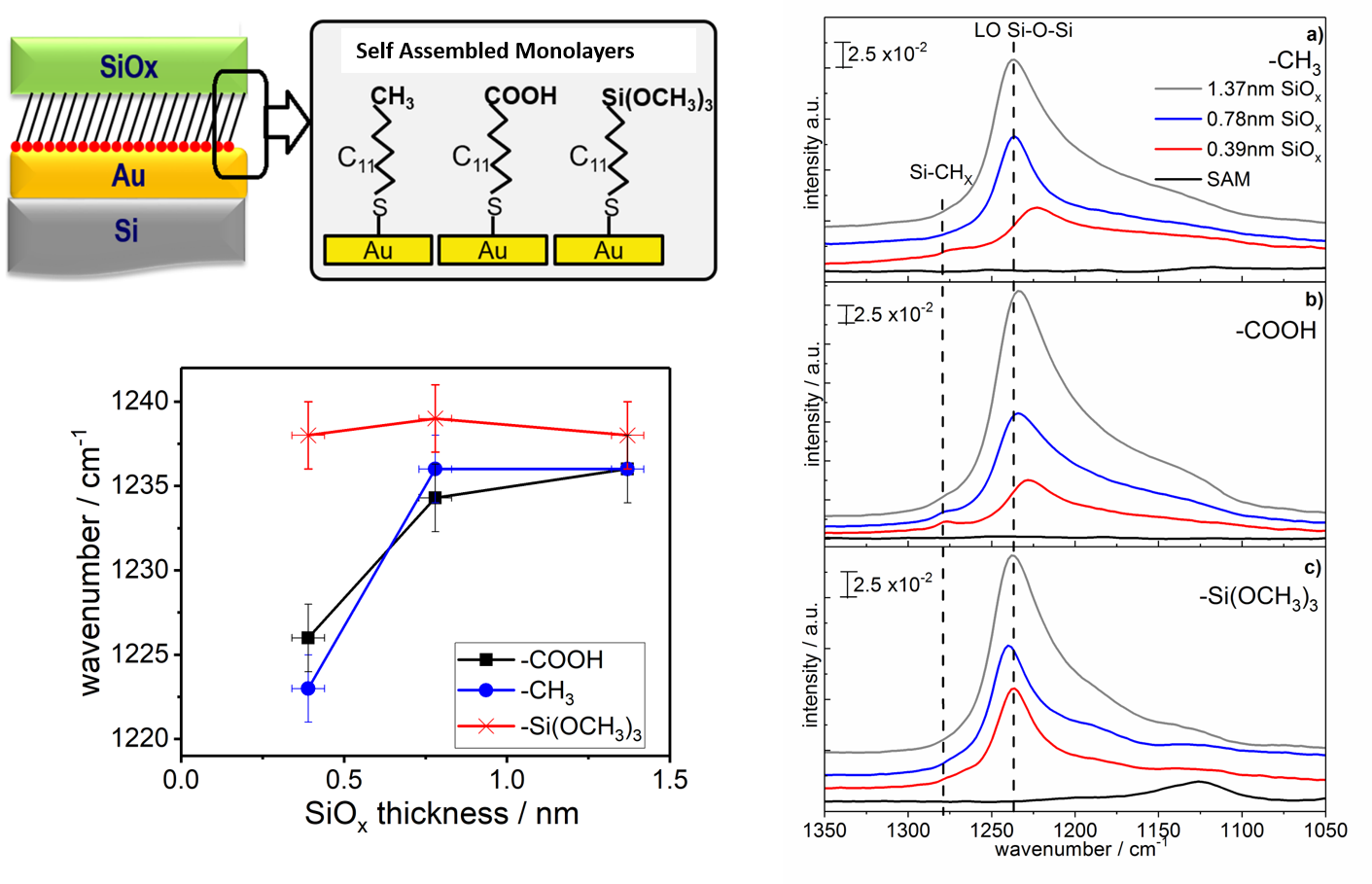}
\end{center}
\caption{Overview of different aspects of nucleation of SiO$_x$ onto SAM-modified surfaces. \textbf{Top left:} schematic of SAMs-modified Au surfaces for the subsequent plasma deposition of SiO$_x$ films. \textbf{Bottom left:} evolution of the Si-O-Si band maximum as function of SAM and thickness of the SiO$_x$ coating. \textbf{Right:} IRRAS spectra showing the Si-O-Si streching band of SiO$_x$ for different SAMs and SiO$_x$ thickness.  Adapted with permission from Hoppe \textit{et al} \cite{Hoppe.2017}. Copyright 2017 IOP Publishing. } \label{fig:grundmeier43}
\end{figure}

A molecular approach to systematically analyze the interaction between a plasma and an organic surface was done by Hoppe \textit{et al}. \cite{Hoppe.2017}. The authors employed self-assembled monolayers on gold for the analysis of plasma-induced surface reactions as a function of the molecular surface termination. This approach, in combination with the application of infrared reflection absorption spectroscopy allowed for the spectroscopy analysis of the surface reaction in the outermost surface region. Different organothiol self-assembled monolayer were deposited, with terminating groups ranging from methyl to trimethoxysilane \cite{Hoppe.2017}. Ultra-thin SiO$_x$ films with thickness values ranging from 0.4 to 1.4 nm were deposited on top by a microwave plasma. Infrared spectra (see figure \ref{fig:grundmeier43}) confirmed that aliphatic terminating groups such as -CH$_{3}$ or -COOH are easily destroyed by the impinging atomic oxygen resulting in high defect densities and impeded cross-linking during early stages of SiO$_x$ film growth. This is due to etching of the polymer's aliphatic fragments which are incorporated into the SiO$_x$ network thereby leading to a poor barrier performance of the total layer.  On the other hand, the trimethoxysilane group protected the aliphatic chain of the SAM and led to a reduced defect density of the ultra-thin SiO$_x$-film. The authors could explain this effect by the reaction of the trimethoxysilane with the oxygen containing plasma leading to a Si–O terminated surface on which the SiO$_x$ film was formed. Barrier properties for layers deposited on SAMs with -Si(OCH$_3$)$_3$ terminating groups could therefore be maintained due to the formation of low defect, highly cross-linked SiO$_x$ layer on top. 

\textbf{Specific aspects of selected polymers:} As just seen, the substrate composition strongly influences the nucleation and growth of PECVD and PEALD thin films. Generally, the interfacial interactions and bonding between the polymeric substrate and the deposited thin films are determined by the substrates surface energy, the functional surface groups and chemical composition as well as topography. As discussed in the prior chapters, plasma pretreatment is a viable tool to alter these surface properties, providing two opposing mechanisms (etching and deposition).
 
 In the initial stages, etching and therefore ablation dominate the deposition. For polyolefin surfaces, such as PE and PE-like surfaces, breaking of the polymer chains occur during the etching.\cite{Jaritz_2017} Therefore, in order to  ensure uniform thin film growth on PP usually an intermediate SiOCH coating is applied prior to deposition of SiO$_x$. The carbonyl rich interlayers bond well to the polyolefin surface and the SiO$_x$ layer. 

Another approach to improve adhesion is the modification of PP with PDMS. PDMS migrates to the surface of the PP during processing, i.e., injection molding, and increases the silicon content of the surface. As in the case of the -Si(OCH$_3$)$_3$ terminated groups seen by Hoppe \textit{et al} \cite{Hoppe.2017}  etching and oxidation of the PDMS during plasma deposition results in formation of SiO$_x$ on the surface of the modified PP. The SiO$_x$ fulfills the two roles of protecting the underlying PP from etching, and to provide a chemical surface to which the silicon organic fragments from the plasma can subsequently bind.\cite{Hoppe.2018}. 

Since technical used polyolefins contain additives, their effect on thin film growth has to be considered. For example, surface penetrating anti-bloc particles are introduced in plastic foil to prevent fusing of the foil during coiling. These particles can compromise the defect free growth of a thin film \cite{Behm_2016}. Also, degradation of e. g., added polyamide containing lubricant can result in weak boundary layers and therefore compromise the integrity of the polymer/coating interface.\cite{behm2013plasma} 

For PET the surface roughness is a key factor for film growth. While slight roughening during the thin film deposition, respectively pretreatment, has no effect on the adhesion of the barrier coating, overtreating the surface by e.g., oxygen or argon plasma results in a roughness which prevents the formation of a closed thin film. Due to an overly rough surface the growth of the thin film is distorted an results in incongruent protrusions within the thin film and a loss in barrier performance.\cite{Bahre_2013}

\begin{figure}[htb]
\begin{center}
\includegraphics[width=0.3\textwidth]{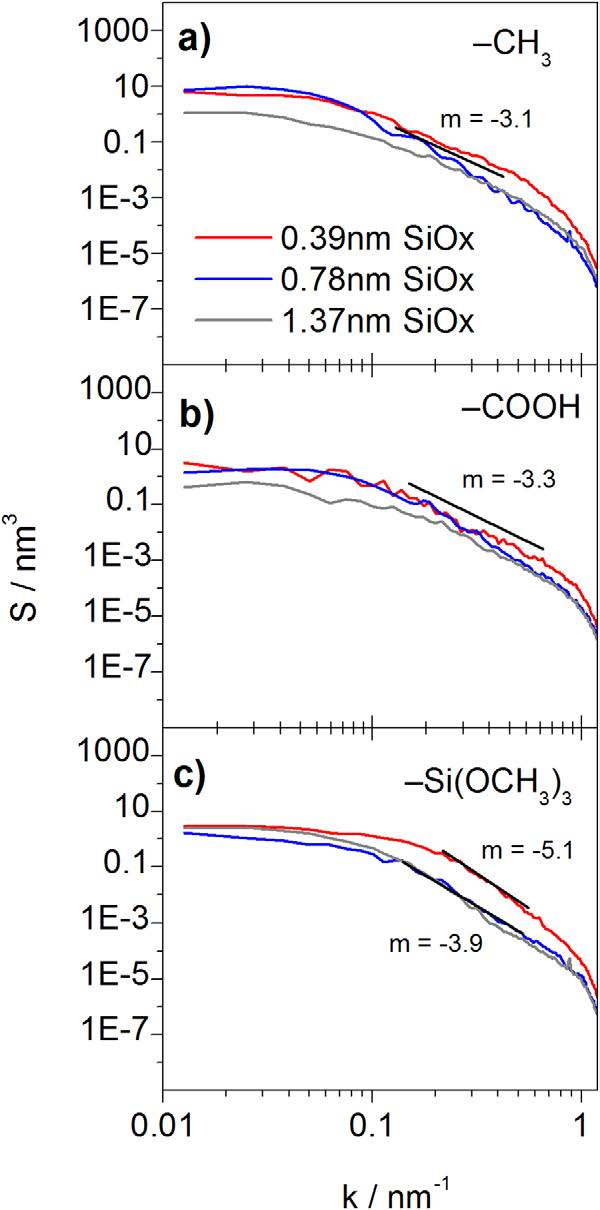}
\end{center}
\caption{Structure factor curves taken from AFM images of the SiO$_x$ films already shown in figure \ref{fig:grundmeier43}. The substrates are Au-coated Si wafers terminated with differently capped SAMs: \textbf{(a)} Undecanethiol \textbf{(b)} 11-Mercapto-1-undecanoic acid \textbf{(c)} 11-Mercapto-1-
undecyltrimethoxysilane. Adapted with permission from Hoppe \textit{et al} \cite{Hoppe.2017}. Copyright 2017 IOP Publishing.
\label{fig:GrundmeierHoppe8}}
\end{figure}

Interestingly, molecular surface chemistry can also affect the development of surface roughness during film growth as shown for of model substrates with very low roughness. Figure \ref{fig:GrundmeierHoppe8} shows the structure factor curves obtained from the analysis of AFM images of SiO$_x$ films deposited on PVD gold films covered with different SAMs, which were already described in figure \ref{fig:grundmeier43}. Although the overall shape of the curves is similar, a clear difference is observed in the intermediate scaling regime, marked in the figure by the slope \textit{m}. While SiO$_x$ films deposited onto aliphatic surfaces show a  behaviour coherent with a diffusion driven growth mechanism (\textit{m} around 3), SiO$_x$ grown onto -Si(OCH$_3$)$_3$ groups shows behaviour associated to super-roughening (\textit{m} around 5), deviating significantly from the diffusion-controlled growth of the other model chemistries.\cite{Hoppe.2017} 

\subsection{Combination of PECVD and PEALD processes}\label{subsec43}


In the field of gas barrier layers, low permeation rates of oxygen or water vapor are desired. To achieve these targets, alternative approaches for suppressing OTR of polymer substrates have been adopted taking advantage of PECVD and PEALD processing of materials. As reported by Kirchheim \textit{et al}. \cite{Kirchheim.2018} and Mitschker \textit{et al}. \cite{Mitschker.2017}, defect formation during growth of such layers is one of the key factors determining their barrier performance. In comparison to coatings obtained from PECVD processes, PEALD showed significantly lower defect densities leading to improved barrier performances as reported by Mitschker \textit{et al}. \cite{Mitschker.2017}. However, as outlined in previous sections \ref{subsec41} and \ref{subsec42} deposition rates from PECVD exceed those from PEALD therefore making PEALD an expensive alternative to PECVD. 

\begin{figure}[h]
    \centering
    \begin{subfigure}[b]{0.24\textwidth}
        \centering
        \includegraphics[width=\textwidth]{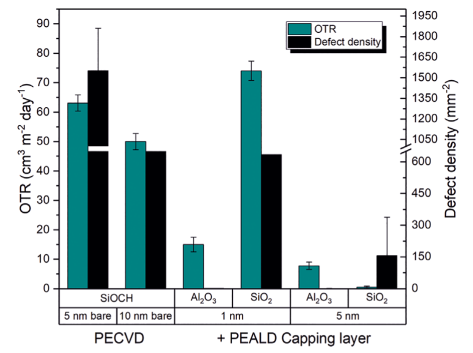}
        \label{fig:Capping_SiOCH}
    \end{subfigure}
    \hfill
    \begin{subfigure}[b]{0.24\textwidth}
        \centering
        \includegraphics[width=\textwidth]{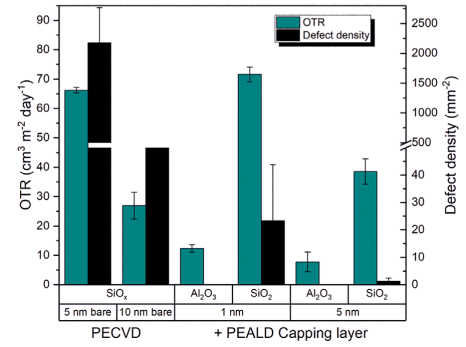}
        \label{fig:Capping_SiOx}
    \end{subfigure}
\caption{Oxygen transmission rate (OTR) and defect density of capping experiments. \textbf{Left:} Capping layers grown on PECVD SiOCH coatings. \textbf{Right:} capping layers grown on PECVD SiO$_x$ coatings. Adapted with permission from \cite{Gebhard.2017b}. Copyright 2017 John Wiley and Sons. }
\label{fig:CappingResults}
\end{figure}

\textbf{Seeding/capping approach:}
A novel method, comprising the advantages of both techniques, was proposed by Gebhard \textit{et al}. \cite{Gebhard.2017b}. Thin layers of up to 5\,nm thickness deposited from PEALD processes were combined with PECVD grown layers deposited either a) before the PEALD layer (capping route) or b) after the PELAD layer (seeding route). As PEALD layers, SiO$_2$ and Al$_2$O$_3$ were used in varying thickness ranging from 1\,nm to 5\,nm. As PECVD layers, SiOCH and SiO$_x$ coatings were deposited with thicknesses from 5\,nm to 15\,nm. For the capping route, SiO$_2$ and Al$_2$O$_3$ PEALD layers were grown on either 5\,nm SiOCH or 5\,nm SiO$_x$ films. As can be seen from figure \ref{fig:CappingResults} Al$_2$O$_3$ capping layer exhibited almost no defects irrespective of their thickness or their underlying PECVD coating. With regards to barrier performance an improvement by a factor of 8 was obtained for the Al$_2$O$_3$ capping. When capping with SiO$_2$, defect densities were higher than for Al$_2$O$_3$, but still significantly reduced compared to initial SiOCH and SiO$_x$ coatings. Using SiO$_x$ as the underlying PECVD coating proved to be beneficial for obtaining low defect densities, whereas the best barrier properites were obtained with SiOCH as the underlying PECVD coating. In this composition, an improvement in barrier performance by two orders of magnitude was reported. 

\begin{figure}[htb]
\begin{center}
\includegraphics[width=0.4\textwidth]{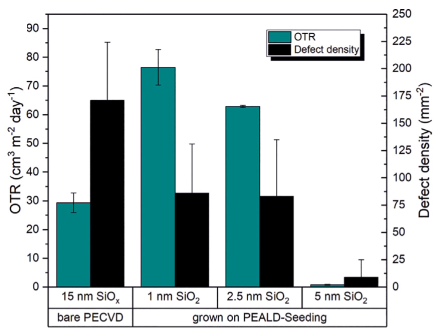}
\end{center}
\caption{Oxygen transmission rate (OTR) and defect density of 15\,nm PECVD SiO$_x$ grown on SiO$_2$ seeding layers. Adapted with permission from \cite{Gebhard.2017b}. Copyright 2017 John Wiley and Sons.}
\label{fig:Seeding}
\end{figure}

For the seeding route, 15\,nm SiO$_x$ grown by PECVD was deposited on SiO$_2$ PEALD layers of varying thickness. The obtained improvement was similar to the one from the capping approach with SiO$_2$ grown on PECVD SiOCH. The defect density was significantly reduced and the barrier performance improved by two orders of magnitude. as can be seen from figure \ref{fig:Seeding}.

\begin{figure}[h]
    \centering
    \begin{subfigure}[b]{0.24\textwidth}
        \centering
        \includegraphics[width=\textwidth]{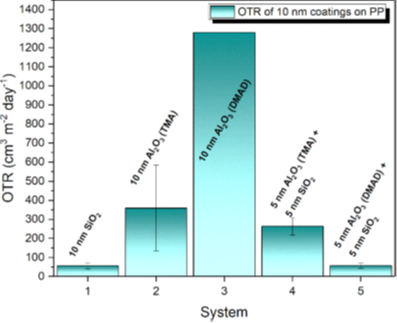}
        \label{fig:Dyads_OTR}
    \end{subfigure}
    \hfill
    \begin{subfigure}[b]{0.24\textwidth}
        \centering
        \includegraphics[width=\textwidth]{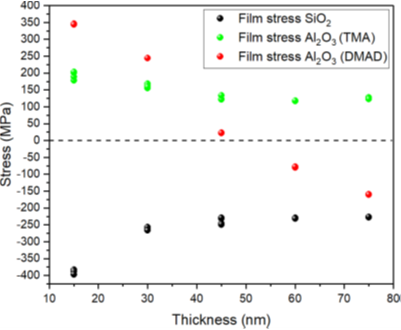}
        \label{fig:Dyads_Stress}
    \end{subfigure}
\caption{\textbf{Left:} OTRs of different combinations of PEALD grown layers deposited on 23\,$\mathrm{\mu}$m PP foil. \textbf{Right:} dependence of film stress of SiO$_2$ and Al$_2$O$_3$ on thin film thickness. Adapted with permission from \cite{Gebhard.2017a}. Copyright 2018 American Chemical Society.}
\label{fig:DyadResults}
\end{figure}

\textbf{Dyads:}
Inorganic binary oxides and nitrides such as SiO$_2$, TiO$_2$, Al$_2$O$_3$, Si$_3$N$_4$ grown via plasma assisted ALD based processes have been successfully demonstrated to improve the gas barrier performance of polymers \cite{Gebhard.2016, Adringa.2015, Hoffmann.2015}. Nano and microdefects in gas barrier layers (GBLs) can affect the transport mechanisms and the material properties have a distinct influence on the performance of the GBLs. Various combinations of inorganic oxides (for e.g., Al$_2$O$_3$/TiO$_2$, Al$_2$O$_3$/HfO$_2$) have also been investigated and the formation of such nanolaminates were found to improve the barrier performance of the employed polymer substrates. However, the investigated nanolaminates were of higher thickness ($\geq$ 50\,nm) and the chosen substrates are rather of higher thicknesses ($\geq$ 100\,$\mathrm{\mu}$m) for specific applications such GBLs in advanced electronics for encapsulation. PEALD can yield abrupt interfaces between different materials and the substrates. By varying the sequence of material deposited, it is possible to obtain nanolaminates (dyads) of amorphous and dense materials at low temperatures through which the grain boundary formation can be suppressed. Grain boundaries can be potential diffusion pathways for gases to penetrate through. 
Gebhard \textit{et al}. investigated dyads of very thin inorganic layers grown via PEALD by the sequential deposition of two different material systems, namely Al$_2$O$_3$ and SiO$_2$ \cite{Gebhard.2017a}. The comparative study on the stacking of Al$_2$O$_3$ and SiO$_2$ dyads of various thickness’ on PP via PEALD using different precursor combinations for Al namely trimethyl aluminium (TMA) vs the modified Al precursor [3-(Dimethylamino)propyl]-dimethyl Aluminum, (DMAD) was performed. For SiO$_2$ the standard Bis(diethylamino)silane (BDEAS) precursor was used. The two combinations for dyads, using either TMA or DMAD for the deposition of Al$_2$O$_3$, were compared to the respective binary oxide compounds. By performing in-situ quartz crystal microbalance (QCM) experiments using specifically designed QCM crystals with a spin-coated polypropylene (PP) top layer, the growth of the inorganic single oxides was monitored in dependence of an oxygen plasma pretreatment for the two different Al precursors. The mechanical properties in terms of residual stress and barrier performance were also studied. Al$_2$O$_3$ thin films from DMAD of 10\,nm thickness exhibited no gas barrier when applied on biaxial oriented PP, while Al$_2$O$_3$ from TMA improved the OTR by a factor of 3.6 (fig. \ref{fig:DyadResults}). For SiO$_2$, of the same thickness, an improvement of barrier performance by a factor or 23.3 was found. As for TMA, QCM hinted towards a more diffusive interface as etching during the first cycles need to be taken into consideration. Giving the findings from in-situ QCM experiments, a more diffusive interface could be beneficial in terms of improved barrier performance which could be explained by a better sticking of the inorganic thin film on the substrate, due a gradual increase of the inorganic component. The role of the interface had a distinct influence on the barrier performance. Dyads of 15\,nm SiO$_2$ + 15\,nm Al$_2$O$_3$ revealed residual stresses of -114 and +113\,MPa for the two different Al precursors used (fig. \ref{fig:DyadResults}). These findings were found to be encouraging as the barrier performance seems to be dependent to a major degree on the SiO$_2$ coating while the Al$_2$O$_3$ coating can be used to tailor the residual stress of the GBL materials.


\section{Outlook/perspectives}\label{sec5}


\textbf{In-situ/operando analysis:}
As shown in the articles many spectroscopic methods have been already developed for the understanding of plasma/substrate interactions. However, also in this regard the transfer from vacuum to atmospheric plasmas creates need to adapt analytical methods. Moreover, the coupling of different complementary methods such as FTIR-spectroscopy and mass spectroscopy or optical emission spectroscopy could be further developed. Of particular interest  is the extension of traditionally ultra high vacuum characterization techniques to high pressure ranges. For example, in the case of ambient pressure X-ray photoelectron spectroscopy (AP-XPS), technical improvements in the design of X-ray sources and analyzers has lead to an important rise in the number of laboratory based setups in recent years. The possibility to investigate surface chemistry in-situ and under realistic pressure conditions has led to a explosion of research in the field of electrochemical interfaces \cite{Wi.2022}, catalysis \cite{Han.2021,Nguyen.2019}, or reactions under humidity conditions \cite{BLUHM201071} to put only a few examples. The combination of AP-XPS, mass spectrometry and reflection IR spectroscopy, both in laboratory and synchrotron facilities is particularly interesting, due to the complementary nature of the techniques and the potential to provide comprehensive information about surface reactions \cite{Zhu:ok5039}. 

\textbf{Simulation of plasma/surface interactions:}
Given the multitude of different kind of discharges, species, and surfaces relevant to PECVD, only a small subset of plasma/surface interactions has been studied and even these not comprehensively. Applied methods such as molecular dynamics (e.g., classical, reactive, density functional theory based) cannot resolve the intrinsic time and length scales of all states of matter involved  (i.e., plasma, gas, solid).~\cite{Neyts2017} A burden which is common to almost all plasma applications. For the sputtering and the deposition of metals and metal nitrides in Ar and Ar/N$_2$ plasmas, a series of machine learning surrogate models for temporal scale-bridging has recently been proposed.~\cite{kruger_machine_2019, gergs_efficient_2022, gergs_physics-separating_2022, gergs_physics-separating_2023} Plasma/surface interactions and their effect on the deposited thin film have been predicted with high physical fidelity for experimental process times of up to 45 minutes within 34 GPU hours (as compared to more than approximately 8 million CPU years).~\cite{gergs_physics-separating_2023} The derived concept can readily be transferred to PECVD, PE-ALD, and other plasma applications (e.g., plasma-enhanced catalysis) and is particularly well suited to handle the high dimensional parameter space inherent to PECVD due to the feasibility of the proposed randomized data generation and the profound generalization capabilities of the data-driven modeling approach.

\textbf{Process developments:}
Beside the advanced understanding of plasma processes under reduced pressure, atmospheric plasma processes (APP) are of increasing importance for polymer surface activation and thin film deposition. Reasons are simplified equipment and process integration as well as lower costs. Moreover, many developments consider the design of plasma jets enabling localized surface modification.
However, some challenges remain such as the low plasma density or large gas volumes required to sustain the process. PEALD-inline processes are of increasing importance for the integration of ALD in production lines. For high process rates, precursors and plasma can be spatially separated and continuously fed to the substrate surfaces moving through different zones. This approach is known as spatial PE-ALD. Spatial PE-ALD in comparison to temporal ALD and enables high throughput values. In addition, it can be performed at atmospheric pressure.

\textbf{Thin film deposition on recycled polymers:}
The European Commission's Plastics Strategy, published in January 2018, sets the goal of recycling at least half of plastic waste by 2030 and thus also substantially increasing the proportion of recycled materials used in plastic products. However, molecular contaminations in such post-consumer recyclates (PCR) could migrate from the packaging into the food - a processes which has to be prevented.  Thin PECVD layers could act as efficient migration barriers for contaminants. 
However, the plasma activation and plasma thin film deposition on such complex polymeric substrates has been rarely addressed \cite{Amberg.2022,Prestes.2015}. Future research should consider the interaction between such complex polymer surfaces with activating and polymerizing plasmas, as well as the migration processes of contaminants in the system consisting of plastic, the PECVD layer and a foodstuff. It is likely that especially for such substrates the combined plasma and surface analytical approach is imperative.


\subsection*{Acknowledgments} Funded by the Deutsche Forschungsgemeinschaft (DFG, German Research Foundation) – Project-ID 138690629 – TRR 87. 

\subsection*{Data Availability Statement}
The data for this work are available upon reasonable request to the author.

\bibliography{wileyNJD-AMA}

\end{document}